\documentclass[12pt]{article}

\usepackage[reqno]{amsmath}
\usepackage{mathrsfs}
\usepackage{amssymb}

\usepackage{bbm}
\usepackage{epsfig}
\usepackage{array}
\usepackage{float}
\usepackage{color}
\usepackage{graphicx}

\usepackage{a4}
\usepackage{a4wide}
\usepackage{wasysym}
\usepackage{url}

\def\gs{\mathrel{
   \rlap{\raise 0.511ex \hbox{$>$}}{\lower 0.511ex \hbox{$\sim$}}}}
\def\ls{\mathrel{
   \rlap{\raise 0.511ex \hbox{$<$}}{\lower 0.511ex \hbox{$\sim$}}}}

\newcommand{\be}{\begin{equation}}
\newcommand{\ee}{\end{equation}}
\newcommand{\bena}{\begin{eqnarray}}
\newcommand{\eena}{\end{eqnarray}}

\newcommand{\sla}[1]{#1 \hspace*{-1ex}/}

\begin{document}

\begin{titlepage}
\title{\vspace*{-2.0cm}
\bf\Large
Enhancing Dark Matter Annihilation into Neutrinos
\\[5mm]\ }

\author{
Manfred Lindner\thanks{email: \tt manfred$.$lindner@mpi-hd.mpg.de}~~,~~
Alexander Merle\thanks{email: \tt alexander$.$merle@mpi-hd.mpg.de}~~,~~
Viviana Niro\thanks{email: \tt viviana$.$niro@mpi-hd.mpg.de}
\\ \\
{\normalsize \it Max-Planck-Institut f\"ur Kernphysik,}\\
{\normalsize \it Postfach 10 39 80, 69029 Heidelberg, Germany}
}
\date{\today}
\maketitle
\thispagestyle{empty}

\begin{abstract}
\noindent
We perform a detailed and quasi model-independent analysis of direct annihilation of Dark Matter into neutrinos. Considering different cases for scalar and fermionic Dark Matter, we identify several settings in which this annihilation is enhanced, contrary to some statements in the literature. They key point is that several restrictions of, e.g., a supersymmetric framework do not hold in general. The mass generation mechanism of the neutrinos plays an important role, too. We illustrate our considerations by two examples that are not (as usually) suppressed by the smallness of the neutrino mass, for which we also present a numerical analysis. Our results can be easily used as guidelines for model building.
\end{abstract}

\end{titlepage}

\section{Introduction}
\label{sec:intro}

Astrophysics and cosmology provides us with compelling evidences of a form of non-luminous and non-baryonic matter, the so-called Dark Matter (DM), which should account for almost 23\% of the total energy density of the Universe. One possibility to detect DM is to search for neutrinos coming from annihilations of DM particles in the Milky Way galactic center and in the galactic halo, in dwarf spheroidal galaxies, and in celestial bodies, like the Earth and the Sun. For recent studies on this subject see, e.g., Refs.~\cite{Hooper:2008cf,Feng:2008qn,Kumar:2009ws,Niro:2009mw}. The main scope of this paper is to analyze the DM pair-annihilation into neutrino final states as model-independent as possible. The most interesting signal to look for at neutrino telescopes is represented by a monochromatic neutrino signal, which can be produced by DM pair-annihilations directly into $\nu \bar{\nu}$ (or into $\nu \nu$ ($\bar{\nu}\bar{\nu}$), if we allow for lepton number violating (LNV) processes). The neutrino energy spectra produced in these annihilation channels are constituted of a soft part (originating from higher order processes) and a sharp line at an energy $E_\nu \simeq m_\chi$. We restrict our study to the two-body direct production, since this is the golden channel for DM discovery at neutrino telescopes. In our analysis, we distinguish between Dirac and Majorana neutrinos and, in the latter case, we consider different neutrino mass generation mechanisms, since they affect the relation between the physical neutrino mass and the neutrino Yukawa couplings.

In the literature, it is often stated that direct annihilation into neutrinos is always suppressed. Though this statement is perfectly true in supersymmetric models (see, e.g., Refs.~\cite{Hooper:2008cf,Goldberg:1983nd,Jungman:1995df}), it does not hold in general. The key point of our paper is to investigate the cases, where such a suppression is not present. This can already be seen in a simple see-saw framework (using only one generation of $\nu_L$ and $N_R$): Here, it is often correctly argued that, even though the neutrino Dirac Yukawa coupling $y_D$ might be sizable, it still does not allow for a sizable direct production of two (light) neutrinos, since it couples only to one doublet $\nu_L$ as well as to one singlet interaction eigenstate $N_R$. Even though the doublet state practically consists of the light mass eigenstate only, $\nu_L\approx \nu'_L$, the fraction of light mass eigenstates within the singlet $N_R$ is suppressed by a tiny mixing angle $\theta\sim \frac{m_D}{M_R}$, which reduces the corresponding production rate. However, when going to a see-saw type II framework, the situation changes: For illustration, one could add a mass term $m_L \overline{(\nu_L)^C} \nu_L$, whose mass is given by the scalar triplet vacuum expectation value (VEV) $v_T$ times the corresponding Yukawa coupling $y_T$. The VEV $v_T$ is forced to be small, due to the correction to the $\rho$-paremeter, which allows $y_T$ to be larger. But $y_T$, in turn, will be related to the annihilation rate of DM into neutrinos in an $s$-channel diagram, in case the DM can couple to the Higgs triplet. Furthermore, the left-handed mass $m_L$ is related to the physical neutrino mass by $m_\nu=m_L-\frac{m_D^2}{M_R}$ (see Sec.~\ref{sec:Neutrino:mass:see-saw} for details), which implies that a partial cancellation in the see-saw formula might be responsible for the correct neutrino mass and allows for an $m_L$ that could be much larger than $m_\nu$. Another example is if the triplet does not obtain a VEV at all: Then, there will be no restriction on the size of $y_T$ from the neutrino sector and the annihilation rates could be even larger. Note that, however, in this case other bounds may apply, e.g.\ from limits on lepton flavour violation~\cite{Kakizaki:2003jk} or neutrino-less double beta decay~\cite{Petcov:2009zr}. This shows how the suppression of the annihilation rates into neutrinos is not always realized, and we will analyze all possibilities in a model-independent way.

This paper is structured as follows: After a brief review of the possible neutrino mass terms in Sec.\,\ref{sec:Neutrino:mass}, we report in Sec.\,\ref{sec:nuprod} the various possibilities for monoenergetic neutrino production, considering explicitly scalar and fermionic DM, as well as the corresponding $s,t,$ and $u$ channels. For simplicity, we do not extend the Standard Model (SM) gauge group, but we contemplate different $SU(2)_L$ representations for the DM and the mediator particles. This kind of systematic analysis was not presented before in the literature in the context of DM annihilation into neutrinos (see Ref.~\cite{Agrawal:2010fh} for the classification of DM scattering off nucleons dominated by spin-dependent interactions). Our results are summarized in Sec.\,\ref{sec:general}, before we give a discussion of examples of unsuppressed scenarios in Sec.\,\ref{sec:Unsuppressed}. We explicitly show the behavior of the annihilation cross sections for a promising $s$-channel and $t$-channel diagram, considering both, the case of scalar and fermionic DM. Our results are then compared to experimental limits on $\mu$ and $\tau$ decays, as well as on lepton flavour violating (LFV) processes. The constraints coming from neutrino searches are also considered. All cross sections we have used are listed in Appendix~A. For specific models in which the DM particles annihilate mainly in neutrinos see, e.g., Refs.\,\cite{Falkowski:2009yz,Blennow:2009ag}.

\section{Neutrino mass terms}
\label{sec:Neutrino:mass}

Throughout our work we consider the SM gauge group, $SU(3)_{c}\times SU(2)_{L}\times U(1)_{Y}$. In this framework, the left-handed components of the neutrinos and the charged leptons form doublets under $SU(2)_L$, while the right-handed components of the neutrinos, if present, will be total singlets:
\be
L_{\alpha\,L}=
\begin{pmatrix}
\nu_{\alpha} \\ \alpha
\end{pmatrix}_{L}\sim ({\bf 1},{\bf 2},-1)\,,
\qquad \nu_{\alpha\,R}\sim ({\bf 1},{\bf 1},0)\,,
\ee
where $\alpha$ is the generation index ($\alpha=e,\mu,\tau$). Depending on the nature of the neutrinos, different mass terms can be present in the Lagrangian. For more details on the physics of massive neutrinos, we refer to Refs.\,\cite{Akhmedov:1999uz,Grimus:2003es,Barger:2003qi,Mohapatra:2005wg,Giunti:2007ry}.

\subsection{Dirac mass term}
\label{sec:Neutrino:mass:Dirac}

If the neutrinos are Dirac particles, they will get their mass only by SM-like Yukawa couplings:
\be
\mathcal{L}_{\rm Yukawa}=-Y^{\alpha \beta}_D \,\overline{L_{\alpha L}}\,\tilde{H}\,\nu_{\beta R} + h.c.
\stackrel{\langle H \rangle = v_H}{\longrightarrow} \mathcal{L}_{\rm mass}=-v_H Y_D^{\alpha \beta} \overline{\nu_{\alpha L}} \nu_{\beta R} + h.c.\,,
\label{eq:Dirac_1}
\ee
where $H$ is the SM Higgs, $v_H=174$~GeV is its VEV, and $\tilde{H}\equiv i \sigma_2 H^*$. The neutrino mass matrix $M_\nu=v_H Y_D$ in the flavour basis is then related to the diagonal neutrino mass matrix $D_\nu={\rm diag}(m_1,m_2,m_3)$ by
\be
M_\nu= U D_\nu U^\dagger\,,
\label{eq:basis_Dirac}
\ee
where $U$ is the leptonic mixing matrix. Note that, since the neutrino masses are small, the Yukawa couplings in Eq.~\eqref{eq:Dirac_1} must be tiny, of the order of $Y_D\sim 10^{-12}$.

\subsection{Majorana mass term}
\label{sec:Neutrino:mass:Maj}

If neutrinos are Majorana particles, terms of the form $\overline{(\nu_L)^C} \nu_L$ or $\overline{(\nu_R)^C} \nu_R$ are present. In the first case, for example a scalar triplet field $T$, with
\bena
T=
\begin{pmatrix}
T^-/\sqrt{2} & T^{0}\\
T^{--} & -T^-/\sqrt{2}
\end{pmatrix}\,,
\eena
would explain a term that is gauge invariant under the SM gauge group:
\be
\mathcal{L}_{\rm Yukawa}=-\frac{1}{2}\,
Y^{\alpha \beta}_L\,\overline{(L_{\alpha L})^C}\,(i \sigma_2\, T)\,L_{\beta L} + h.c.
\;\rightarrow \;\mathcal{L}_{\rm mass}=-\frac{1}{2}m_L^{\alpha \beta} \overline{(\nu_{\alpha L})^C} \nu_{\beta L} + h.c.\,,
\label{eq:Maj_L}
\ee
where $m_L^{\alpha \beta}=v_T Y_L^{\alpha \beta}$, with $v_T$ being the VEV of the neutral component of the scalar triplet.

If right-handed neutrinos are present, explicit Majorana mass terms are possible,
\be
\mathcal{L}_{\rm mass}=-\frac{1}{2}\,
M^{\alpha \beta}_R\,\overline{(\nu_{\alpha R})^{C}}\,\nu_{\beta R} + h.c.
\label{eq:Maj_R}
\ee
Such a Majorana mass term could arise from some high energy embedding, whose symmetries might be broken at the grand unification scale. In this case, we would expect $M_R$ to be of the order of $10^{14}-10^{16}$~GeV. In any case, since $M_R$ is not related to the VEV $v_H$, there is no reason for it to be at the electroweak scale.

\subsection{See-saw mechanisms}
\label{sec:Neutrino:mass:see-saw}

Combining Dirac and Majorana mass terms leads to the different see-saw mechanisms, where the smallness of the neutrino mass is a consequence of the heavy right-handed neutrino fields. For a type I see-saw mechanism one needs a Dirac mass terms as well as a pure Majorana mass term for right-handed neutrinos (denoted $N_R$ in this context). The complete neutrino mass term, after electroweak symmetry breaking, is given by
\bena
\mathcal{L}_{\rm mass}&=&- \overline{\nu_L} m_D N_R - \frac{1}{2}
\overline{(N_R)^C} M_R N_R + h.c. =\\\nonumber
&=&-(\overline{\nu_L},\overline{(N_R)^C})
 \left(\begin{array}{c|c}
 0     & m_D\\ \hline
 m_D^T & M_R\\
 \end{array}\right)
 \begin{pmatrix}
 (\nu_L)^C\\
 N_R
 \end{pmatrix},
 \label{eq:see-sawI_1}
\eena
where $m_D=v_H Y_D$ is the Dirac mass matrix and $M_R$ is the Majorana mass matrix for the right-handed neutrinos. The former is connected to the electroweak scale $v_H$, while the latter can have a much larger value. Bringing the above mass matrix into a block-diagonal form yields
\be
\mathcal{L}_{\rm mass}\approx -(\overline{\nu'_L},\overline{(N'_R)^C})
 \left(\begin{array}{c|c}
 -m_D M_R^{-1} m_D^T & 0\\ \hline
 0 & M_R\\
 \end{array}\right)
 \begin{pmatrix}
 (\nu'_L)^C\\
 N'_R
 \end{pmatrix},
 \label{eq:see-sawI_2}
\ee
where the rotated states are denoted by $\nu'$ and $N'$, and we have neglected the small corrections to the heavy neutrino masses. The rotation required is only a very tiny one and the corresponding mixing angles between heavy and light states are or order $\theta\approx \frac{m_D}{M_R}\sim 10^{-14}-10^{-12}$. As a consequence, the interaction eigenstate $\nu_L$ is essentially a light mass eigenstate, while $N_R$ has only a small fraction $\frac{m_D}{M_R}$ of the light mass eigenstate.

In the flavour basis, the light neutrino mass matrix is given by
\be
 M_\nu\equiv -m_D M_R^{-1} m_D^T = U D_\nu U^T\,,
 \label{eq:basis_Majorana}
\ee
where $U$ is the ordinary leptonic mixing matrix.

If beyond the terms of Eq.\,\eqref{eq:see-sawI_1} also a left-handed Majorana mass term $m_L$ is present, a type II see-saw mechanism will be induced, which leads to a light neutrino mass matrix of the form
\be
 M_\nu\equiv m_L-m_D M_R^{-1} m_D^T,
\ee
where $m_L=v_T Y_L$. The corresponding rotation angles are approximately given by
\be
 \theta\simeq \frac{m_D}{M_R-m_L} \approx \frac{m_D}{M_R}\,,
 \label{eq:see-sawII_2}
\ee
where we have used the fact that $m_L \ll M_R$, since the correction to the $\rho$-parameter forces the triplet VEV $v_T$ to be $\lesssim \mathcal{O}(1{\rm~GeV})$. The diagonalization in this case is analogous to the one in Eq.~\eqref{eq:basis_Majorana}.

\section{Classification of the possibilities for neutrino production}
\label{sec:nuprod}

Let us first present the general logic of our analysis: Starting from a quantum field theory point of view, we have different possibilities to assign representations of the SM gauge group to the different particles involved. Depending on the gauge quantum numbers assigned to the DM particle and to the neutrino, specific annihilation processes will be allowed in the $s,t,u$ channels. We restrict our model-independent analysis to the cases where the DM particle $\chi$ and the mediator particle $\phi$ are in a singlet, doublet, or triplet representation of $SU(2)_L$.\footnote{A generalization to more complicated cases is straightforward.} In general, for a scalar $\psi_s$ and for a fermion $\psi_f$ (where $\psi=\chi,\phi$) we use the following conventions:
\bena
\psi_{s;1}\sim({\bf 1},{\bf 1},0)\,,& &
\psi_{f;1}\sim({\bf 1},{\bf 1},0)\,,\nonumber\\
\psi_{s;2}=
\begin{pmatrix}
\psi^+ \\
\psi^0
\end{pmatrix}\sim({\bf 1},{\bf 2},1)\,,& &
\psi_{f;2}=
\begin{pmatrix}
\psi^0 \\
\psi^-
\end{pmatrix}\sim({\bf 1}, {\bf 2},-1)\,, \nonumber\\
\psi_{s;3}=
\begin{pmatrix}
\psi^+/\sqrt{2} & \psi^{++}\\
\psi^0 & -\psi^+/\sqrt{2}
\end{pmatrix}\sim({\bf 1}, {\bf 3},2)\,,& &
\psi_{f;3}=
\begin{pmatrix}
\psi^-/\sqrt{2} & \psi^{0}\\
\psi^{--} & -\psi^-/\sqrt{2}
\end{pmatrix}\sim({\bf 1}, {\bf 3},-2)\,.\nonumber
\eena
We will comment later on the possibility of having an $SU(2)_L$ triplet with zero hypercharge.

Throughout our analysis we will consider the limit $m_\nu\rightarrow 0$ with, however, $m_\nu \ne 0$. This means that, even though the neutrino mass can be neglected in all kinematical considerations, it will still be important to know the mechanism that generates this mass. As explained in the illustrative example in Sec.~\ref{sec:intro}, a different type of neutrino mass can in many cases also lead to different restrictions on certain couplings, leading to very different results in some cases (see Sec.~\ref{sec:summary}).

We present an analysis of all the possible production channels in a way as model-independent as possible, extending the work presented in Ref.\,\cite{Beltran:2008xg}, in which the authors restricted themselves to the case of Dirac DM annihilating through an $s$ channel diagram. In Secs.\,\ref{sec:S} and\,\ref{sec:F}, we present the results for direct neutrino production in the cases of scalar and fermionic DM, respectively. To be exhaustive, we explicitly divide the results into four different neutrino scenarios: Dirac neutrinos, for which the left-handend and the right-handed neutrinos are both present and independent; Majorana neutrinos, in which case the singlet neutrinos $\nu_{\alpha R}$ are not present and the right-handed neutrinos are simply given by $(\nu_{\alpha L})^{C}$; Majorana neutrinos with see-saw type I or type II, if the right-handed neutrinos $\nu_{\alpha R}$ are present and acquire a general mass.

In our study, we do not consider explicitly the case of vector DM. It is known, indeed, that spin one DM particles can have a sizable branching ratio into neutrinos, see, e.g., Ref.\,\cite{Blennow:2009ag}. The main goal of our analysis is, instead, to show that also in the framework of scalar or fermionic DM the direct neutrino production can be relevant. However, for completeness, we report in Appendix~A the explicit expressions for the annihilation cross sections in the case of vector DM, too.

We wish to recall that the neutrino production through DM annihilations into three body final states has also been vastly discussed in the literature. For instance, in Ref.\,\cite{Bell:2008ey} the authors analyzed the electroweak bremsstrahlung processes $\chi\chi\rightarrow \nu \bar{\nu}Z$ and $\chi\chi\rightarrow\nu e W$. The hadronic decays of the weak bosons can lead to the production of photons, which can then be used to further constrain the annihilation cross section value, see, e.g., Ref.\,\cite{Kachelriess:2007aj} for the $Z$-strahlung process. Moreover, the DM annihilation into neutrinos will induce, at loop level, electromagnetic final states, for which the synchrotron radiation bounds of Ref.\,\cite{Hooper:2008zg} can be imposed; see Ref.\,\cite{Dent:2008qy} for an exhaustive discussion on this aspect.

\subsection{Scalar Dark Matter}
\label{sec:S}
%
This Section summarizes the results we have obtained for the case of scalar DM, considering singlet, doublet, and triplet representations of $SU(2)_L$. The basic assumptions are that the scalar DM has a null VEV, $\langle \chi \rangle=0$, and that it is stable, for example because being odd under some $Z_2$-parity while all the other SM particles are even.

\subsubsection*{\noindent Scalar mediator, $s$-channel}
In the case of a singlet, doublet, or triplet scalar mediator, the following Yukawa interactions with the neutrinos are allowed:
\bena
\mathcal{L}_{Y_{\nu;1}}&=&
-Y^{\alpha \beta}_{\nu;1}\,
\overline{(\nu_{\alpha R})^{C}}\,\phi_{s;1}\,\nu_{\beta R}\,+\,h.c.\,,\\
\mathcal{L}_{Y_{\nu;2}}&=&
-Y^{\alpha \beta}_{\nu;2}\,
\overline{L_{\alpha L}}\,\tilde{\phi}_{s;2}\,\nu_{\beta R}\,+\,h.c.\,,\\
\mathcal{L}_{Y_{\nu;3}}&=&
-Y^{\alpha \beta}_{\nu;3}\,
\overline{(L_{\alpha L})^{C}}\,(i\sigma_{2}\,\phi_{s;3})\,L_{\beta L}\,
+\,h.c.\,,
\label{eq:tripletNu}
\eena
where $\alpha$ and $\beta$ are flavour indices. We have defined $\tilde{\phi}\equiv i \sigma_2 \phi^{*}$, with $\sigma_{2}$ being the second Pauli matrix. Note that the entries of the Yukawa coupling matrices are in general complex numbers, and that a triplet scalar mediator with zero hypercharge, $\phi_{s;3}\sim({\bf 1}, {\bf 3},0)$, does not couple to ordinary neutrinos in an $s$ channel diagram.

The singlet scalar mediator $\phi_{s;1}$ can couple to a pair of scalar DM particles, which transform as a singlet, doublet, or triplet under $SU(2)_L$. However, it will always produce a physical right-handed (light) neutrino as well as a left-handed (light) anti-neutrino. Both these particles are sterile, and making them interacting would require a coupling to the Higgs field (or, equivalently, a helicity flip), which is proportional to $m_\nu$. This would lead to a negligible muon flux at neutrino telescopes. Notice also that the coupling $\overline{(\nu_R)^{C}}\,\phi_{s;1}\,\nu_R$ could, in general, generate a violation of lepton number $L$ and is hence connected to Majorana neutrinos. Indeed, the coupling of the singlet scalar $\phi_{s;1}$ to the two singlet neutrinos either directly violates lepton number or it forces the singlet scalar to carry lepton number.
In the latter case, the coupling of $\phi_{s;1}$ to the SM Higgs, $H^\dag H \phi_{s;1}$, will be problematic. However, if such a coupling is forbidden in certain specific models, one might still be able to conserve lepton number.

If the doublet scalar mediator $\phi_{s;2}$ does not get a VEV, the entries in the Yukawa coupling matrix $Y_{\nu;2}$ can be large, as they do not contribute to the neutrino mass. However, a fundamental problem arises from the coupling to the scalar DM: Since we consider only the cases for which the scalar DM particle does not get a VEV, the corresponding vertex must arise from a fundamental 3-scalar coupling in the Higgs potential. In $SU(2)_L$, such a fundamental 3-scalar coupling is impossible, since we have
${\bf 2} \otimes {\bf 1} \otimes {\bf 1}={\bf 2}$,
${\bf 2} \otimes {\bf 2} \otimes {\bf 2}={\bf 2} \oplus {\bf 2} \oplus {\bf 4}$,
${\bf 2} \otimes {\bf 3} \otimes {\bf 3}={\bf 2} \oplus {\bf 2} \oplus {\bf 4} \oplus {\bf 4} \oplus {\bf 6}$.
This problem could be overcome if one allowed for a non-vanishing VEV $\langle\phi_{s;2}\rangle \ne 0$. However, in this way the Yukawa coupling $Y_{\nu;2}$ becomes directly proportional to the light neutrino mass for the case of Dirac neutrinos. In the presence of a see-saw situation, the Yukawa coupling could in principle be sizable, since it is not directly related to the neutrino mass. In spite of that, since the light mass eigenstate of $\nu_R$ must be produced, this possibility is suppressed by the mixing angle $\theta$ between the heavy and light neutrinos. This will be of $\mathcal{O}\left(\frac{m_D}{M_R}\right)$, and hence very small for the standard value of $M_R\sim \mathcal{O}(10^{16} \text{GeV})$.

The interaction of the triplet scalar mediator $\phi_{s;3}$ with neutrinos, Eq.\,\eqref{eq:tripletNu}, in general induces a violation of lepton number (in analogy to the singlet scalar mediator $\phi_{s;1}$) and is thus associated with Majorana neutrinos and not with Dirac neutrinos. In case the scalar mediator has a null VEV, the neutrino coupling $Y_{\nu;3}$ will be unsuppressed, since is not constrained by the neutrino mass scale. Furthermore, two active neutrinos are produced, since the triplet scalar couples to $\overline{(\nu_L)^C} \nu_L$. This conclusion does not depend on the particular neutrino mass model considered. Indeed, in the case of see-saw type I, the correction factor resulting from $\nu_L$ being not an exact mass eigenstate is given by $(1-\theta)^2\simeq 1$. If the neutrinos acquire a mass through a see-saw type II model, the only difference is the presence of an {\it additional} Higgs triplet with VEV, in order to have the correct see-saw type II neutrino mass formula.

The DM vertex for the case $\langle\phi_{s;3}\rangle=0$ can come from a fundamental 3-scalar term in the Higgs potential. This coupling will be allowed only if $\chi$ is an $SU(2)_L$ doublet. In this case, the important term in the Lagrangian will be of the form
\bena
\mathcal{L}^{(2,3)}_{\chi\phi}&\supset&
\gamma^{(2,3)}_{\chi\phi}\,(\chi^{\dag}_{s;2}\phi_{s;3}\tilde{\chi}_{s;2})\,+\, h.c.
\label{eq:trilinear}
\eena

If the triplet scalar mediator has a nonzero VEV, $\langle\phi_{s;3}\rangle\ne0$, it will contribute to a Majorana neutrino mass term proportional to $\overline{(\nu_L)^C} \nu_L$ and it will induce a see-saw type II situation. Thus, the light neutrino mass matrix would be given by
\begin{equation}
 M_\nu=v_T Y_{\nu;3}-v_H^2 Y_{\nu;2} M_R^{-1} Y_{\nu;2}^T\,,
 \label{eq:TvSee-sawII_3}
\end{equation}
where $v_H$ is the electroweak VEV and $v_T$ is the triplet scalar VEV. To yield physically realistic light neutrino masses, the entries in the Yukawa coupling matrix $Y_{\nu;3}$ of the triplet to the neutrinos must be very small, in case the triplet contribution dominates the physical neutrino masses.\footnote{Note that, although the VEV is forced by the correction to the $\rho$-parameter only to be $v_T\lesssim \mathcal{O}(1~{\rm GeV})$, the Yukawa coupling still needs to be tiny to yield sub-eV neutrino masses.} On the other hand, the combination of the Dirac Yukawa coupling $Y_{\nu;2}$ and the heavy neutrino mass matrix $M_R$ has to be tiny as well, if this part dominates the physical neutrino mass.

The only case where we can have larger values for $Y_{\nu;3}$, which, in turn, could lead to larger annihilation rates, is the one where there is a cancellation between $v_T Y_{\nu;3}$ and $v_H^2 Y_{\nu;2} M_R^{-1} Y_{\nu;2}^T$ in Eq.\,\eqref{eq:TvSee-sawII_3}. For simultaneously having Yukawa couplings of $\mathcal{O}(0.1)$ and sub-eV neutrino masses, this cancellation would, however, need to be at the level of $10^{-8}$ (for $v_T\approx 1$~GeV), which would require a strong fine-tuning. Nevertheless, this possibility might be motivated in a specific model.

The corresponding couplings of the $SU(2)_L$ triplet scalar mediator $\phi_{s;3}$ with non-vanishing VEV to the DM particles can arise from the following terms in the Lagrangian:
\bena
\mathcal{L}^{(1,3)}_{\chi\phi}&=&\lambda^{(1,3)}_{\chi\phi}\,
(\chi^{\dag}_{s;1}\chi_{s;1}){\rm Tr}(\phi^{\dag}_{s;3}\phi_{s;3})\,,\\
\mathcal{L}^{(2,3)}_{\chi\phi}&=&\lambda^{(2,3)}_{\chi\phi}\,
(\chi^{\dag}_{s;2}\chi_{s;2}){\rm Tr}(\phi^{\dag}_{s;3}\phi_{s;3})\,+\,
\beta^{(2,3)}_{\chi\phi}\,
(\chi^{\dag}_{s;2}\phi^{\dag}_{s;3}\phi_{s;3}\chi_{s;2})\,+\nonumber\\
&+&\left[\gamma^{(2,3)}_{\chi\phi}\,(\chi^{\dag}_{s;2}\phi_{s;3}\tilde{\chi}_{s;2})\,+\,
h.c.\right],\\
\mathcal{L}^{(3,3)}_{\chi\phi}&=&\lambda^{(3,3)}_{\chi\phi}\,
{\rm Tr}(\chi^{\dag}_{s;3}\chi_{s;3}){\rm Tr}(\phi^{\dag}_{s;3}\phi_{s;3})\,+\nonumber\\
&+&\left[\xi^{(3,3)}_{\chi\phi}\,
{\rm Tr}(\chi_{s;3}\chi_{s;3}){\rm Tr}(\phi^{\dag}_{s;3}\phi^{\dag}_{s;3})\,+\,
h.c.\right]\,.
\eena
The triplet scalar mediator appears as the most promising case for having a sizable neutrino production. However, depending on the specific model, its coupling to the leptons can be subject to constraints coming from different experiments. We postpone the explicit discussion of these bounds to Sec.\,\ref{sec:Unsuppressed:triplet}.

\subsubsection*{\noindent $Z$-boson mediator, $s$-channel}
The coupling between the neutrinos and the $Z$-boson comes from the gauge-kinetic term in the Lagrangian. We define the covariant derivative as
\be
D_{\mu}=\partial_{\mu}
+i\frac{g}{2}(\vec{\sigma} \cdot \vec{W}_{\mu})
+i\frac{g^{\prime}}{2}Y B_{\mu}\,,
\label{eq:cov2}
\ee
with $\vec{\sigma}$ being the Pauli matrices. The couplings $g$ and $g^{\prime}$ are, respectively, the gauge couplings of $SU(2)_L$ and $U(1)_Y$, and $Y$ is the hypercharge of the field that couples to the $Z$-boson. The corresponding gauge fields are denoted by $W_\mu$ and $B_\mu$. Introducing the physical states
\bena
W^{\pm}_{\mu}&=&\frac{W^{1}_{\mu}\mp iW^{2}_{\mu}}{\sqrt{2}}\,,\\
A_{\mu}&=&B_{\mu} \cos \theta_{W}+W^{3}_{\mu}\sin \theta_{W}\,,\\
Z_{\mu}&=&-B_{\mu} \sin \theta_{W}+W^{3}_{\mu}\cos \theta_{W}\,,
\eena
where $\theta_W$ is the Weinberg angle, the interaction term of the neutrinos with the $Z$-boson is given by
\be
\mathcal{L}^{\rm kin}_{L}=\overline{L_{L}}i \gamma^{\mu}D_{\mu}L_{L}
\rightarrow -\frac{g}{2 \cos \theta_{W}}\overline{\nu_{L}}\gamma^{\mu}\nu_{L} Z_{\mu}\,.
\ee
Only if the DM particle transforms as a doublet or triplet under $SU(2)_L$, it can couple to the $Z$-boson. The specific couplings arise from the following gauge-kinetic terms:
\bena
\mathcal{L}^{\rm kin}_{\chi;2}&=&(D_{\mu}\chi_{s;2})^{\dag}(D^{\mu}\chi_{s;2})\nonumber\\
&\rightarrow& -\frac{i g}{2 \cos \theta_{W}}(\cos^{2}\theta_{W}+Y\sin^{2}\theta_{W})\,
(\partial_{\mu} \chi^0)^* \chi^{0} Z^{\mu}\,+\,h.c.\,,\\
\mathcal{L}^{\rm kin}_{\chi;3}&=&{\rm Tr}\left[(D_{\mu}\chi_{s;3})^{\dag}(D^{\mu}\chi_{s;3})\right]\,\nonumber\\
&\rightarrow& -\frac{i g}{2 \cos \theta_{W}}(2 \cos^{2}\theta_{W}+Y\sin^{2}\theta_{W})\,
(\partial_{\mu} \chi^0)^* \chi^{0} Z^{\mu}\,+\,h.c.\,,
\eena
where the covariant derivative for $\chi_{s;2}$ is defined analogously to Eq.\eqref{eq:cov2} (with $Y$ being the hypercharge of the DM), while for $\chi_{s;3}$ it is given by
\be
D_{\mu}\chi_{s;3}=\partial_{\mu}\,\chi_{s;3}+
i\frac{g}{2}\left[\vec{\sigma} \cdot \vec{W}_{\mu},\chi_{s;3}\right]+
i\frac{g^{\prime}}{2}Y B_{\mu}\chi_{s;3}\,.
\ee

\subsubsection*{\noindent Fermionic mediator, $t$ and $u$-channels}
For the $t$ and $u$-channel diagrams, either the scalar DM or the fermionic mediator has to be flavoured, such as in the case of a sneutrino DM~\cite{Arina:2007tm} or a neutrino mediator~\cite{Haber:1984rc}. This property has to be taken into account in any specific model and it will decide about the actual existence of a $t$-channel diagram. For definiteness, throughout our discussion, we suppose that the scalar DM particle carries a flavour. Our conclusions are as well applicable to the case in which the fermionic mediator is flavoured.

We consider a fermionic mediator, whose left and right components can transform under $SU(2)_L$ as singlets, doublets or triplets. If the fermionic mediator is an $SU(2)_L$ singlet, $\left[\phi_{f;1}\right]_{L,R}$, the following interaction terms will be allowed:
\bena
\mathcal{L}^{(1,1)}_{\chi \phi \nu}=&\,
\mathcal{T}^{(1,1)}_{\alpha k}\,
\overline{\nu^\alpha_{R}}\,\chi^k_{s;1}
\left[\phi_{f;1}\right]_{L}\,+\,h.c.\,,
\eena
\bena
\mathcal{L}^{(2,1)}_{\chi \phi \nu}=&\,
\mathcal{T}^{(2,1)}_{\alpha k}\,
\overline{L^\alpha_{L}}\,\tilde{\chi}^k_{s;2}\left[\phi_{f;1}\right]_{R}\,+\,h.c.\,,
\label{eq:tScalarDM}
\eena
where $\mathcal{T}^{(i,j)}_{\alpha k}$ are trilinear couplings, with $\alpha$ being an index in flavour space and $k$ being the index that denotes the lightest scalar particle. In general, indeed, different flavoured states of the scalar particle $\chi^\beta_{s;1}$ can exist. The DM particle will then be identified as the lightest particle among the mass eigenstates, $\chi^k_{s;1}=W_{k \beta}\chi^\beta_{s;1}$, with $W$ being a rotation matrix. The indices $(i,j)$ are, respectively, the $SU(2)_L$ representations of the DM and of the fermionic mediator. If the DM is a singlet scalar, it will only couple to sterile neutrinos, while if it is the neutral component of a doublet, active neutrinos can be produced.

If the fermionic mediator is an $SU(2)_L$ doublet, $\left[\phi_{f;2}\right]_{L,R}$, the interaction terms that lead to a coupling between the DM particle and the neutrino will be:
\be
\begin{split}
\mathcal{L}^{(1,2)}_{\chi \phi \nu}=&
\,\mathcal{T}^{(1,2)}_{\alpha k}\,
\overline{L^\alpha_{L}}
\left[\phi_{f;2}\right]_{R}\chi^k_{s;1}\,+\,h.c.\,,\\
\mathcal{L}^{(2,2)}_{\chi \phi \nu}=&
\,\mathcal{T}^{(2,2)}_{\alpha k}\,
\overline{\nu^\alpha_{R}}
(i \sigma_{2}\chi^k_{s;2})^{T}\left[\phi_{f;2}\right]_{L}\,+\,h.c.\,,\\
\mathcal{L}^{(3,2)}_{\chi \phi \nu}=&
\,\mathcal{T}^{(3,2)}_{\alpha k}\,
\overline{(L^\alpha_{L})^C}
(i \sigma_{2} \chi^k_{s;3})\left[\phi_{f;2}\right]_{L}\,+\,h.c.
\end{split}
\ee
In this case, a singlet and a triplet scalar DM can couple to active neutrinos, while only sterile neutrinos will be produced if the DM is an $SU(2)_L$ doublet.

Finally, if the fermionic mediator is an $SU(2)_L$ triplet, $\left[\phi_{f;3}\right]_{L,R}$, we can have the following couplings:
\be
\begin{split}
\mathcal{L}^{(2,3)}_{\chi \phi \nu}=&\,
\mathcal{T}^{(2,3)}_{\alpha k}\,
\overline{L^\alpha_{L}}
\left[\phi_{f;3}\right]_{R}\chi^k_{s;2}\,+\,h.c.\,,\\
\mathcal{L}^{(3,3)}_{\chi \phi \nu}=&\,
\mathcal{T}^{(3,3)}_{\alpha k}\, 
{\rm Tr}\left\{\overline{\nu^\alpha_{R}}
\left[\phi_{f;3}\right]_{L}\chi^k_{s;3}\right\}\,+\,h.c.
\end{split}
\ee
Active neutrinos arise from a doublet scalar DM, while a triplet scalar DM couples only to sterile neutrinos. Moreover, if the fermion mediator is an $SU(2)_L$ triplet with $Y=0$, it can also couple to a scalar DM triplet with $Y=0$ and to a right-handed neutrino. This coupling would, however, produce only sterile neutrinos and thus lead to a negligible flux.

As in the case of scalar DM pair-annihilations into neutrinos through a scalar exchange, the couplings involved in the $t$-channel process are subject to experimental limits, coming in particular from LFV processes. For example, if active neutrinos are produced and if the fermionic mediator belongs to a doublet or a triplet representation of $SU(2)_L$, the couplings involved in the $t$-channel diagram will also contribute to the $\mu \to e \gamma$ decay. Another experimental constraint that could be present is the one on the actual existence of the fermion particle mediating the process. We will comment on these points in Sec.\,\ref{sec:Unsuppressed:singlet}.

\subsection{Fermionic Dark Matter}
\label{sec:F}

In this Section we consider the DM as fermionic particle and, in analogy to the scalar case, we allow for $SU(2)_L$ singlet $\left[\chi_{f;1}\right]_{L,R}\sim({\bf 1}, {\bf 1},0)$, doublet $\left[\chi_{f;2}\right]_{L,R}\sim({\bf 1}, {\bf 2},-1)$, and triplet $\left[\chi_{f;3}\right]_{L,R}\sim({\bf 1}, {\bf 3},-2)$ representations.

\subsubsection*{\noindent Scalar mediator, $s$-channel}
For the $s$-channel, the considerations for the neutrino vertex are exactly the same as in the scalar DM case. Therefore in the following we will focus on the DM vertex only. An intermediate scalar singlet $\phi_{s;1}$ could couple to all types of fermionic DM under consideration:
\bena
\mathcal{L}^{(1,1)}_{Y_{\chi;1}}&=&\,
-Y^{(1,1)}_{\chi;1}\,
\overline{\left[\chi_{f;1}\right]_{L}}
\left[\chi_{f;1}\right]_{R}\phi_{s;1}\,+\,h.c.\,,
\label{eq:first}\\
\mathcal{L}^{(2,2)}_{Y_{\chi;1}}&=&\,
-Y^{(2,2)}_{\chi;1}\,
\overline{\left[\chi_{f;2}\right]_{L}}
\left[\chi_{f;2}\right]_{R}\phi_{s;1}\,+\,h.c.\,,\\
\mathcal{L}^{(3,3)}_{Y_{\chi;1}}&=&\,
-Y^{(3,3)}_{\chi;1}\,
{\rm Tr}\left\{\overline{\left[\chi_{f;3}\right]_{L}}
\left[\chi_{f;3}\right]_{R}\right\}\phi_{s;1}\,+\,h.c.
\eena
In all the above cases, the left and right components of the DM particle belong to the same representation of $SU(2)_L$, i.e., the DM is a vector-like fermion. Note that another possible expression can be obtained by replacing $\overline{\left[\chi_{f}\right]_{L}}$ with $\overline{\left[\chi_{f}\right]_{R}^{\;\;C}}$ in Eq.~\eqref{eq:first}. The interaction term obtained in this way is associated with Majorana DM, since it can induce a Majorana mass term.\footnote{Similar to the neutrino case, we would require a very specific setup to be able to consistently define any global $U(1)$-symmetry that is able to distinguish DM and anti-DM.} As in the scalar DM case, the problem arises at the neutrino vertex, since only sterile neutrinos can be produced by a scalar singlet.

If the scalar mediator is an $SU(2)_L$ doublet, $\phi_{s;2}$, we could have the following couplings to the DM particle:
\bena
\mathcal{L}^{(1,2)}_{Y_{\chi;2}}&=&\,
-Y^{(1,2)}_{\chi;2}\,
\overline{\left[\chi_{f;1}\right]_{L}}
(i \sigma_{2} \phi_{s;2})^{T}
\left[\chi_{f;2}\right]_{R}\,+\,h.c.\,,\\
\mathcal{L}^{(3,2)}_{Y_{\chi;2}}&=&\,
-Y^{(3,2)}_{\chi;2}\,
\phi^{\dag}_{s;2}\overline{\left[\chi_{f;3}\right]_{L}}\,
\left[\chi_{f;2}\right]_{R}\,+\,h.c.
\eena
These possibilities will only be present if the left and right components of the DM particle belong to different representations of $SU(2)_L$, i.e., if the DM is a chiral fermion. As for the scalar DM case, the situation in which $\phi_{s;2}$ has a nonzero VEV can be neglected, since the Yukawa couplings would then have to be proportional to the neutrino mass or a tiny mixing angle $\theta$ between the heavy and light neutrinos would be present.

If the scalar mediator is a triplet under $SU(2)_L$, $\phi_{s;3}$, the following terms are allowed:
\bena
\mathcal{L}^{(2,2)}_{Y_{\chi;3}}&=&
-Y^{(2,2)}_{\chi;3}\,
\overline{\left[\chi_{f;2}\right]^{\;\;C}_{L}}
(i \sigma_{2} \phi_{s;3})\left[\chi_{f;2}\right]_{L}
\,+\,h.c.\,,
\label{eq:last}\\
\mathcal{L}^{(1,3)}_{Y_{\chi;3}}&=&
-Y^{(1,3)}_{\chi;3}\,
{\rm Tr}\left\{\overline{\left[\chi_{f;1}\right]_{L}}
\phi_{s;3}\left[\chi_{f;3}\right]_{R}\right\}
\,+\,h.c.\,,
\eena
where the first term will be present if the DM particle is a vector-like fermion, while the second one will be there if it is a chiral fermion.

An analogous expression to Eq.\,\eqref{eq:last} can be obtained by exchanging the subscripts `$L$' with `$R$' and considering a new Yukawa coupling $Y^{\prime(2,2)}_{\chi;3}$. Notice that, if Eq.~\eqref{eq:last} holds, the triplet scalar mediator in the $s$-channel will be associated only with Majorana DM (as well as Majorana neutrinos), since it leads to terms that violate lepton number. If the scalar triplet acquires a nonzero VEV, a see-saw type II situation will be induced, in analogy to the scalar DM case, to which we refer for more details. Remember that, in principle, also a DM coupling to a triplet scalar with zero hypercharge was possible, but this would not lead to a coupling to SM-like neutrinos.

\subsubsection*{\noindent $Z$-boson mediator, $s$-channel}
A fermionic DM particle can couple to the $Z$-boson if the DM is a doublet or a triplet under $SU(2)_L$. The corresponding couplings arise from the following gauge-kinetic terms in the Lagrangian:
\bena
\mathcal{L}^{\rm kin}_{\chi;2}&=&
\overline{\left[\chi_{f;2}\right]_{L}}
i \gamma^{\mu}D_{\mu} \left[\chi_{f;2}\right]_{L}\,\nonumber\\
&\rightarrow& -\frac{g}{2 \cos \theta_{W}}(\cos^{2}\theta_{W}-Y\sin^{2}\theta_{W})
\overline{\left[\chi_{f;2}\right]^{0}_{L}} \gamma^{\mu} \left[\chi_{f;2}\right]^{0}_{L} Z_{\mu}\,+\,h.c.\,,\\
\mathcal{L}^{\rm kin}_{\chi;3}&=&
{\rm Tr} \left\{
\overline{\left[\chi_{f;3}\right]_{L}}
i \gamma^{\mu}D_{\mu} \left[\chi_{f;3}\right]_{L}\right\}\,\nonumber\\
&\rightarrow& -\frac{g}{2 \cos \theta_{W}}(2\cos^{2}\theta_{W}-Y\sin^{2}\theta_{W})
\overline{\left[\chi_{f;2}\right]^{0}_{L}} \gamma^{\mu} \left[\chi_{f;2}\right]^{0}_{L} Z_{\mu}\,+\,h.c.\,,
\eena
and analogous expressions can be written for the right-handed components $\left[\chi_{f;2}\right]_{R}$ and $\left[\chi_{f;3}\right]_{R}$ of the DM particle. As before, $Y$ denotes the hypercharge of the DM particle.

\subsubsection*{\noindent Scalar mediator, $t$ and $u$-channels}
As in the case of scalar DM, for the $t$ and $u$-channel diagrams, either the fermionic DM or the scalar mediator has to be flavoured, such as in the case of a heavy neutrino DM or a sneutrino scalar mediator. For definiteness, throughout the discussion of this Section we will suppose that the scalar mediator carries a flavour. Our conclusions are as well applicable to the case in which the DM is flavoured.

If the scalar mediator is an $SU(2)_L$ singlet $\phi_{s;1}$, the following interaction terms will be allowed:
\bena
\mathcal{L}^{(2,1)}_{\chi \phi \nu}=&\,
\mathcal{T}^{(2,1)}_{\alpha k}\,
\overline{L^\alpha_{L}}
\left[\chi_{f;2}\right]_{R}\,\phi^k_{s;1}\,+\,h.c.\,,
\label{eq:tFermionicDM}
\eena
\bena
\mathcal{L}^{(1,1)}_{\chi \phi \nu}=&\,
\mathcal{T}^{(1,1)}_{\alpha k}\,
\overline{\nu^\alpha_{R}}
\left[\chi_{f;1}\right]_{L}\,\phi^k_{s;1}\,+\,h.c.\,,
\eena
where $\mathcal{T}^{(i,j)}_{\alpha k}$ are trilinear couplings, with $\alpha$ being an index in flavour space and $k$ being the index that denotes the mass eigenstate of the scalar mediator. The indices $(i,j)$ are, respectively, the $SU(2)_L$ representations of the DM particle and of the scalar mediator. In order not to produce only sterile neutrinos, the right-handed component of the fermionic DM particle has to be an $SU(2)_L$ doublet.

If, instead, the scalar mediator is an $SU(2)_L$ doublet, $\phi_{s;2}$, the interaction terms will be the following:
\be
\begin{split}
\mathcal{L}^{(1,2)}_{\chi \phi \nu}=&\,
\mathcal{T}^{(1,2)}_{\alpha k}\,
\overline{L^\alpha_{L}}
\tilde{\phi}^k_{s;2}\,
\left[\chi_{f;1}\right]_{R}\,+\,h.c.\,,\\
\mathcal{L}^{(2,2)}_{\chi \phi \nu}=&\,
\mathcal{T}^{(2,2)}_{\alpha k}\,
\overline{\nu^\alpha_{R}}\,
(i \sigma_{2}\phi^k_{s;2})^{T}
\left[\chi_{f;2}\right]_{L}\,+\,h.c.\,,\\
\mathcal{L}^{(3,2)}_{\chi \phi \nu}=&\,
\mathcal{T}^{(3,2)}_{\alpha k}\,
\overline{L^\alpha_{L}}
\left[\chi_{f;3}\right]_{R}\,
\phi^k_{s;2}\,+\,h.c.\,,
\end{split}
\ee
among which only the ones involving a singlet or a triplet fermionic DM lead to active neutrinos in the final state. One specific example falling in this category would be a slight extension of the MSSM, with an additional singlet chiral superfield, whose fermionic component acts as DM particle, while the sneutrino is the scalar mediator~\cite{Martin:1997ns}.

Finally, if the scalar mediator is an $SU(2)_L$ triplet, $\phi_{s;3}$, we will have
\be
\begin{split}
\mathcal{L}^{(2,3)}_{\chi \phi \nu}=&\,
\mathcal{T}^{(2,3)}_{\alpha k}\,
\overline{(L^\alpha_L)^C}
(i\sigma_{2}\phi^k_{s;3})\,
\left[\chi_{f;2}\right]_{R}\,+\,h.c.\,,\\
\mathcal{L}^{(3,3)}_{\chi \phi \nu}=&\,
\mathcal{T}^{(3,3)}_{\alpha k}\,
{\rm Tr}\left\{
\overline{\nu^\alpha_{R}}\,\phi^k_{s;3}
\left[\chi_{f;3}\right]_{L}\,\right\}\,+\,h.c.\,,
\end{split}
\ee
where only in the case of a doublet fermionic DM particle the production of active neutrinos is possible.

Note that, in our analysis of the $t$-channel diagram for fermionic DM, we have decided to neglect the possibility that the intermediate scalar mediator acquires a nonzero VEV. In this case, a mixing between the DM particle and the neutrino would be induced. The corresponding constraints on the Yukawa couplings would hence become strongly model-dependent, and general conclusions would not be possible anymore, in contradiction to the aim of this paper. Nevertheless, one can use the guidelines provided here to analyze this case in a certain model, too.

Furthermore, as in the case of scalar DM pair-annihilations, the couplings involved in the $t$-channel process could be subject to the experimental limits coming from LFV processes. We refer to Sec.\,\ref{sec:Unsuppressed:singlet} for more details.

\section{Results}
\label{sec:summary}

\subsection{General Discussion}
\label{sec:general}

\begin{table}[t]
 \centering
 \begin{tabular}{|c|c|r||c|c|c|c|}\hline
 Annihilation & Internal & Dark Matter & Dirac & Majorana & See-saw & See-saw  \\
 channels & mediator & $SU(2)_L$-rep.\ & neutrino & neutrino & type I & type II \\\hline\hline
 $s$ & Scalar {\bf 1} & {\bf 1,\,2,\,3} & $\sla{L}$ & - & $R,\theta^2$ & $R,\theta^2$  \\\hline
 $s$ & Scalar {\bf 3} & {\bf 2} & $\sla{L}$ & \textcolor{red}{$\CheckedBox$} & \textcolor{red}{$\CheckedBox$} & \textcolor{red}{$\CheckedBox$}  \\\hline
 \hline
 $s$ & Scalar {\bf 1}, VEV & {\bf 1,\,2,\,3} & $\sla{L}$ & - & $R,\theta^2$ & $R,\theta^2$  \\\hline
 $s$ & Scalar {\bf 2}, VEV & {\bf 1,\,2,\,3} & $m_\nu$ & - & $\theta$ & $\theta$  \\\hline
 $s$ & Scalar {\bf 3}, VEV & {\bf 1,\,2,\,3} & $\sla{L}$ & $m_\nu$ & - & f.t. \\\hline\hline
$s$ & $Z$-boson & {\bf 2,\,3} & $\XBox (p)$ & $\XBox (p)$ & $\XBox(p)$ & $\XBox(p)$  \\\hline\hline
 $t,\,u$ & Fermion {\bf 1} & {\bf 1} & $R$ & - & $R,\theta^2$ & $R,\theta^2$  \\\hline
 $t,\,u$ & Fermion {\bf 1} & {\bf 2} & \textcolor{red}{$\CheckedBox$} & \textcolor{red}{$\CheckedBox$} & \textcolor{red}{$\CheckedBox$} & \textcolor{red}{$\CheckedBox$}  \\\hline
 $t,\,u$ & Fermion {\bf 2} & {\bf 1,\,3} & \textcolor{red}{$\CheckedBox$} & \textcolor{red}{$\CheckedBox$} & \textcolor{red}{$\CheckedBox$}  & \textcolor{red}{$\CheckedBox$}  \\\hline
 $t,\,u$ & Fermion {\bf 2} & {\bf 2} & $R$ & - & $R,\theta^2$ & $R,\theta^2$  \\\hline
 $t,\,u$ & Fermion {\bf 3} & {\bf 2} & \textcolor{red}{$\CheckedBox$} & \textcolor{red}{$\CheckedBox$} & \textcolor{red}{$\CheckedBox$} & \textcolor{red}{$\CheckedBox$}  \\\hline
 $t,\,u$ & Fermion {\bf 3} & {\bf 3} & $R$ & - & $R,\theta^2$ & $R,\theta^2$  \\\hline
 \end{tabular}
 \caption{\label{tab:summary_scalDM} Scalar Dark Matter cases: \textcolor{red}{$\CheckedBox$} stands for `potentially unsuppressed in at least one channel', $\XBox (p)$ stands for `suppressed for non-relativistic Dark Matter ($p$-wave term)', f.t.\ stands for `fine tuning required between two couplings to get a sizable rate',  $\sla{L}$ stands for `LNV terms are present', - stands for `a see-saw type I and/or type II situation is present', $R$ stands for `yields only right-handed neutrinos', $\theta^n$ stands for `suppressed by the $n$-th power of the mixing angle between heavy and light neutrinos', and $m_\nu$ stands for `the Yukawa coupling involved is proportional to the light neutrino mass'.}
\end{table}

In this Section, we present the results of our analysis. As already discussed in the Introduction, Dark Matter annihilation into neutrinos can be suppressed, e.g., through the proportionality of the corresponding amplitude to the neutrino mass. Another reason for a suppression can come from angular momentum restrictions enforced by the Pauli principle due to the Dark Matter consisting of Majorana fermions. Whenever there is no such reason for a suppression, we will call the corresponding process {\it unsuppressed}.

An overview of our results for the different cases is given in Tables~\ref{tab:summary_scalDM} and~\ref{tab:summary_FermDM}. Note that, although we mark certain cases as `potentially unsuppressed', this might not be true in specific models where additional restrictions apply (e.g., a certain coupling might be forbidden by some symmetry). In this model-independent framework, such peculiarities cannot be taken into account, so when using our results, the reader should always convince himself/herself that they are applicable in the particular case under consideration. Note that by `Majorana neutrinos', we implicitly refer to the case that only left-handed neutrinos exist. In particular, these neutrinos only have a left-handed mass term, see Eq.~\eqref{eq:Maj_L}.

Tables~\ref{tab:summary_scalDM} and~\ref{tab:summary_FermDM} can be used to identify settings in which Dark Matter annihilation into neutrinos may be unsuppressed (as long as there are no other problematic aspects of the model, which might, e.g., constrain one particular coupling to be tiny). These cases can then be used as a guideline for model-building: For example, in Tab.~\ref{tab:summary_scalDM}, one advantageous case can be seen in the second line, where a scalar {\bf 2} is the Dark Matter candidate, which is supplemented by a Higgs triplet that does not get a VEV. A natural framework in which such a setting exists would be a left-right symmetric extension of the scotogenic neutrino mass model~\cite{Adulpravitchai:2009re}, if the left Higgs triplet does not obtain a VEV. Another candidate may be the MSSM, supplemented by right-handed neutrinos that do get a Majorana mass term and by two Higgs triplet superfields without VEV (in this case, one would indeed need both fields for complete cancellation of anomalies). Then, neutralino Dark Matter would just correspond to the third case of Tab.~\ref{tab:summary_FermDM}. Using similar arguments, one can easily construct further settings in which the annihilation rates of Dark Matter into neutrinos are sizable. We will discuss some specific examples in the following.

\begin{table}[H]
 \centering
 \begin{tabular}{|c|c|r||c|c|c|c|}\hline
 Annihilation & Internal & Dark Matter & Dirac    & Majorana & See-saw & See-saw  \\
 channels & mediator & $SU(2)_L$-rep.\ & neutrino & neutrino & type I & type II \\\hline\hline
 $s$ & Scalar {\bf 1} & {\bf 1,\,2,\,3} & $\sla{L}$ & - & $R,\theta^2$ & $R,\theta^2$ \\\hline
 $s$ & Scalar {\bf 2} & {\bf (1,2)\,,(2,3)} & \textcolor{red}{$\CheckedBox$} & - & $\theta$ & $\theta$ \\\hline
 $s$ & Scalar {\bf 3} & {\bf 2$^{M}$} & $\sla{L}$ & \textcolor{red}{$\CheckedBox$} & \textcolor{red}{$\CheckedBox$} & \textcolor{red}{$\CheckedBox$} \\\hline
 $s$ & Scalar {\bf 3} & {\bf (1,3)} & $\sla{L}$ & \textcolor{red}{$\CheckedBox$} & \textcolor{red}{$\CheckedBox$} & \textcolor{red}{$\CheckedBox$} \\\hline\hline
 $s$ & Scalar {\bf 1}, VEV & {\bf 1,\,2,\,3} & $\sla{L}$ & - & $R,\theta^2$ & $R,\theta^2$ \\\hline
 $s$ & Scalar {\bf 2}, VEV & {\bf (1,2)\,,(2,3)} & $m_\nu$ & - & $\theta$ & $\theta$ \\\hline
 $s$ & Scalar {\bf 3}, VEV & {\bf 2$^{M}$} & $\sla{L}$ & $m_\nu$ & - & f.t.\\\hline
 $s$ & Scalar {\bf 3}, VEV & {\bf (1,3)} & $\sla{L}$ & $m_\nu$ & - & f.t.\\\hline\hline
 $s$ & $Z$-boson & {\bf 2,\,3} & \textcolor{red}{$\CheckedBox$}/$\XBox (p)$ & \textcolor{red}{$\CheckedBox$}/$\XBox (p)$ & \textcolor{red}{$\CheckedBox$}/$\XBox (p)$ & \textcolor{red}{$\CheckedBox$}/$\XBox (p)$ \\\hline\hline
 $t\,(u)$ & Scalar {\bf 1} & {\bf 1} & $R$ & - & $R,\theta^2$ & $R,\theta^2$ \\\hline
 $t\,(u)$ & Scalar {\bf 1} & {\bf 2} & \textcolor{red}{$\CheckedBox$} & \textcolor{red}{$\CheckedBox$} & \textcolor{red}{$\CheckedBox$} & \textcolor{red}{$\CheckedBox$} \\\hline
 $t\,(u)$ & Scalar {\bf 1} & {\bf (1,2)} & \textcolor{red}{$\CheckedBox$} & - & $\theta$ & $\theta$ \\\hline
 $t\,(u)$ & Scalar {\bf 2} & {\bf 1,\,3} & \textcolor{red}{$\CheckedBox$} & \textcolor{red}{$\CheckedBox$} & \textcolor{red}{$\CheckedBox$} & \textcolor{red}{$\CheckedBox$} \\\hline
 $t\,(u)$ & Scalar {\bf 2} & {\bf 2} & $R$ & - & $R,\theta^2$ & $R,\theta^2$ \\\hline
 $t\,(u)$ & Scalar {\bf 2} & {\bf (1,2)} & \textcolor{red}{$\CheckedBox$} & - & $\theta$ & $\theta$ \\\hline
 $t\,(u)$ & Scalar {\bf 2} & {\bf (1,3)} & \textcolor{red}{$\CheckedBox$} & \textcolor{red}{$\CheckedBox$} & \textcolor{red}{$\CheckedBox$} & \textcolor{red}{$\CheckedBox$} \\\hline
 $t\,(u)$ & Scalar {\bf 3} & {\bf 2} & \textcolor{red}{$\CheckedBox$} & \textcolor{red}{$\CheckedBox$} & \textcolor{red}{$\CheckedBox$} & \textcolor{red}{$\CheckedBox$} \\\hline
 $t\,(u)$ & Scalar {\bf 3} & {\bf 3} & $R$ & - & $R,\theta^2$ & $R,\theta^2$ \\\hline 
 $t\,(u)$ & Scalar {\bf 3} & {\bf (2,3)} & \textcolor{red}{$\CheckedBox$} & - & $\theta$ & $\theta$ \\\hline
 \end{tabular}
 \caption{\label{tab:summary_FermDM} Chiral and vector-like fermionic Dark Matter cases: \textcolor{red}{$\CheckedBox$} stands for `potentially unsuppressed in at least one channel', $\XBox (p)$ stands for `suppressed for non-relativistic Dark Matter ($p$-wave term)', f.t.\ stands for `fine tuning required between two couplings to get a sizable rate', $\sla{L}$ stands for `LNV terms are present', - stands for `a see-saw type I and/or type II situation is present', $R$ stands for `yields only right-handed neutrinos', $\theta^n$ stands for `suppressed by the $n$-th power of the mixing angle between heavy and light neutrinos', $m_\nu$ stands for `the Yukawa coupling involved is proportional to the light neutrino mass', $x/y$ stands for `$x$ applies for Dirac DM and $y$ applies for Majorana DM, and the superscript ${M}$ stands for `this coupling is present for Majorana DM only'.}
\end{table}

\subsection{Some unsuppressed cases}
\label{sec:Unsuppressed}

For a quantitative statement, one has to calculate the relevant annihilation cross sections. This is made more complicated by the possibility that the final and/or initial fermions could be Majorana particles. For $s$-channel diagrams, this difficulty can be cured easily be applying the effective vertex method, see Ref.~\cite{Kayser:1989iu}. For the $t$- and $u$-channel diagrams, one has to be careful since they cannot be distinguished physically, which tells us that we have to perform a coherent summation over the two amplitudes, and then, in the end, divide the square of the total amplitude by a factor of $2$ to avoid double counting in the final state~\cite{Haber:1984rc}. We have performed these calculations and list in Appendix~A the explicit expressions for the annihilation cross sections for all the different cases. The results are reported in a model-independent way and can therefore be used for any specific model.

To illustrate how to use our results, we will discuss two example cases in this section, one involving an $s$ and one involving a $t$-channel diagram: For a scalar DM particle, the $s$-channel annihilation diagram can be relevant in the presence of a triplet scalar mediator with zero VEV. Moreover, this case will be present only for Majorana neutrinos. The explicit expression of the annihilation cross section can be found using Eq.\,\eqref{eq:S:scalar}. Another promising situation for neutrino production is given by a $t$-channel diagram with a singlet, a doublet, or a triplet fermion exchange. In the first case the DM particle should be a doublet under $SU(2)_L$, in the second case a singlet or a triplet, and in the third case it should be a doublet. For the $t$-channel diagram, the annihilation cross section will be determined mainly by the mass of the mediator, see Eqs.~\eqref{eq:S:t} and~\eqref{eq:S:maj}. For a fermionic DM particle, a triplet scalar exchange in an $s$-channel diagram can give rise to a sizable neutrino production if the left- or right-handed DM particle transforms as a component of a doublet under $SU(2)_L$ and if the neutrinos are Majorana particles. For a chiral fermion DM, instead, the $s$-channel diagram can be relevant if the scalar mediator is a doublet or a triplet under $SU(2)_L$. The first case can be present if the neutrinos are Dirac particles, while the second one can be there if they are Majorana particles. The explicit expression for the annihilation cross section can be found using Eq.~\eqref{eq:F:scalar}. As in the case of scalar DM, another promising case for neutrino production is given by the $t$-channel diagram with a singlet, doublet, or triplet scalar exchange. In the first case, the DM particle should be a doublet under $SU(2)_L$ and in the second case it should be a singlet or a triplet, while in the third case it must be a doublet. Other unsuppressed $t$-channel diagrams might be present if the DM was a chiral fermion, see Sec.~\ref{sec:general}.

For the $t$-channel diagram, the annihilation cross section will be determined mainly by the mass of the DM particle, see Eqs.~\eqref{eq:F:t} and~\eqref{eq:F:maj}. Moreover, if the DM was a Dirac fermion, also the $s$-channel diagram with $Z$-boson exchange could lead to sizable neutrino production. In this case, the annihilation cross section would be proportional to the mass of the DM particle, see Eq.~\eqref{eq:F:Z}. However, particles with strong couplings to the $Z$-boson are constrained by DM direct detection experiments, see Ref.~\cite{Falk:1994es}.

For definiteness, we focus on two different topologies of unsuppressed cases: One involving an $s$-channel diagram, in Sec.~\ref{sec:Unsuppressed:triplet}, and one with a $t$-channel diagram, in Sec.~\ref{sec:Unsuppressed:singlet}. For the first possibility, we consider a triplet scalar exchange with null VEV and a fermionic DM particle that transforms as a doublet under $SU(2)_L$. We explicitly distinguish between the cases of scalar and Majorana fermionic DM. Remember that a triplet scalar exchange in an $s$-channel diagram is associated with Majorana neutrinos only. For the $t$-channel diagram, we also consider a DM particle that is a doublet under $SU(2)_L$. In the context of a scalar DM, we focus on the possibility of a Majorana singlet mediator, while in the case of Majorana DM we consider a scalar singlet mediator. As an example, we consider the case of Majorana neutrinos for the $t$-channel diagrams.

\subsubsection{$s$-channel: The triplet scalar mediator}
\label{sec:Unsuppressed:triplet}

The couplings involved in an $s$-channel diagram with a triplet scalar exchange will not be connected to the neutrino mass, if the triplet has a null VEV. However, the entries of the Yukawa coupling matrix $Y_{\nu;3}$ are constrained by different experimental results, in particular by the limits on $\mu$ and $\tau$ decays, and by the values of the electron and the muon anomalous magnetic moments. In the following, we summarize these bounds.

\subsubsection*{\noindent Experimental constraints}
The singly charged triplet component $\phi^{-}_{s;3}$ might transmit a lepton number violating muon decay with one $\mu^-$-$\overline{\nu}_\mu$-$\phi_{s;3}^-$ and one $e^-$-$\overline{\nu}_e$-$\phi_{s;3}^-$ vertex. Considering the experimental uncertainty on $G_F$ of about $10^{-10}$~GeV$^{-2}$~\cite{Amsler:2008zzb}, obtained through $\mu$-decay measurements, the corresponding diagonal entries of $Y_{\nu;3}$ are set to be:
\be
|Y_{\nu;3}^{ee}|^2\,|Y_{\nu;3}^{\mu\mu}|^2 \lesssim 0.1
\left( \frac{10^{-10} m^2_{\phi}}{\text{GeV}^2}\right)^2\,.
\label{eq:diagMM}
\ee
In general, also the electrically neutral component $\phi^{0}_{s;3}$ of the Higgs triplet will mediate $\mu$-decay. However, the corresponding diagram involves the LFV coupling $Y^{\mu e}_{\nu;3}$ that is constrained much stronger by the experimental limit on the branching ratio for $\mu \to 3 e$ (see later).

The singly charged triplet component $\phi^{-}_{s;3}$ might transmit a lepton number violating $\tau$-decay with one $\tau^-$-$\overline{\nu}_\tau$-$\phi_{s;3}^-$ and one $e^-$-$\overline{\nu}_e$-$\phi_{s;3}^-$ or $\mu^-$-$\overline{\nu}_\mu$-$\phi_{s;3}^-$ vertex. Therefore, the diagonal elements of $Y_{\nu;3}$ receive bounds also from the experimental limit on the $\tau$ lifetime. Taking into account that the uncertainty on $\Gamma_\tau$ is roughly 0.1\%~\cite{Amsler:2008zzb}, we find
\be
|Y_{\nu;3}^{\tau \tau}|^2\,\left(
|Y_{\nu;3}^{ee}|^2\,+\,|Y_{\nu;3}^{\mu\mu}|^2 \right) \lesssim 0.1
\left( \frac{10^{-5} m^2_{\phi}}{\text{GeV}^2}\right)^2\,.
\label{eq:diagTT}
\ee

If the Yukawa coupling matrix $Y_{\nu;3}$ contains off-diagonal terms, the triplet will also have LFV couplings. In this case, the strongest constraint arises from $\mu \to 3e$ decay. Indeed, this process can be mediated at tree-level by the doubly charged component of the triplet $\phi^{--}_{s;3}$ with one $\mu^-$-$e^{+}$-$\phi_{s;3}^{--}$ and one $e^-$-$e^-$-$\phi_{s;3}^{--}$ vertex. From the experiment SINDRUM~I~\cite{Bellgardt:1987du}, we know that the branching ratio $BR(\mu \to 3 e) \lesssim 10^{-12}$ at 90\% confidence level. Therefore, the bound on the off-diagonal entries reads
\be
|Y_{\nu;3}^{e \mu}|^2\,|Y_{\nu;3}^{ee}|^2 \lesssim 5.4
\left( \frac{10^{-11} m^2_{\phi}}{\text{GeV}^2}\right)^2\,.
\ee
Note that the $\mu \to e\gamma$ process naturally arises only at 1-loop level and is therefore suppressed with respect to the $\mu \to 3e$ decay. The branching ratio of the $\tau$ decay into three leptons $l$ (with $l=e,\mu$) is, instead, constrained from the BELLE experiment~\cite{Abe:2007ev} to be $BR(\tau \to lll) \lesssim (2-4)\cdot 10^{-8}$ at 90\% confidence level. This implies the following limit on the off-diagonal $\tau$ Yukawa entries:
\be
|Y_{\nu;3}^{l \tau}|^2\,|Y_{\nu;3}^{ll}|^2 \lesssim 0.6
,\left( \frac{10^{-9} m^2_{\phi}}{\text{GeV}^2}\right)^2\,.
\label{eq:offdiag}
\ee
The Yukawa entries $Y_{\nu;3}^{ee}$ and $Y_{\nu;3}^{\mu \mu}$ are also subject to constraints coming from measurements of the electron and the muon anomalous magnetic moments \cite{Fayet:2007ua}:
\bena
|Y_{\nu;3}^{ee}| \lesssim \mathcal{O}(10^{-4})\,\left(\frac{m_\phi}{{\rm MeV}}\right)\,,\\
|Y_{\nu;3}^{\mu \mu}| \lesssim \mathcal{O}(10^{-6})\,\left(\frac{m_\phi}{{\rm MeV}}\right)\,.
\eena
If we suppose that the diagonal elements $Y_{\nu;3}^{ee}$ and $Y_{\nu;3}^{\mu\mu}$ are of the same magnitude, Eqs.\,\eqref{eq:diagMM} and\,\eqref{eq:diagTT} imply that the only sizable diagonal Yukawa entry is given by the element $Y_{\nu;3}^{\tau \tau}$:
\be
|Y_{\nu;3}^{\tau \tau}|^2 \lesssim {\rm min}\left(1,\,\frac{10^{-1} m^2_\phi}{{\rm GeV}^2}\right)\,,
\ee
where we have explicitly imposed that the Yukawa coupling is at most of order one. Since in our numerical analysis we always consider $m_\phi \gtrsim 100$ GeV, we have $|Y_{\nu;3}^{\tau \tau}|^2 \lesssim 1$. For simplicity, we neglect the contributions coming from the off-diagonal terms of the Yukawa matrix $Y_{\nu;3}$.

In the case of scalar DM, the coupling between the DM particles and the scalar triplet mediator $\phi_{s;3}$ in the $s$-channel arises from a trilinear term in the potential, see Eq.\,\eqref{eq:trilinear}. The existence of this coupling and at the same time the possibility for the scalar triplet to have a null VEV will depend on the actual form of the scalar potential. In a particular model, one has to check that these two conditions are fulfilled.

In the case of vector-like fermionic DM, the coupling between the DM particles and the scalar triplet mediator $\phi_{s;3}$ in the $s$-channel can arise from two different Yukawa couplings: $Y^{(2,2)}_{\chi;3}$, which is related to the DM left-handed components, and $Y^{\prime(2,2)}_{\chi;3}$, which is connected to the DM right-handed components. In case these two couplings result to be of the same order, the $s$-wave contribution to the annihilation cross section will vanish, see Eq.\,\eqref{eq:F:scalar}. However, there is no a priori reason for them to be of the same magnitude. Therefore, we will suppose in our analysis that one of the Yukawa couplings, $Y^{(2,2)}_{\chi;3}$, dominates over the other one, $Y^{\prime(2,2)}_{\chi;3}$.

We want to stress that, even though the scalar triplet can be associated also with a chiral DM (see Tab.~\ref{tab:summary_FermDM}), we neglect this possibility, since strong bounds from electroweak precision measurements apply on new chiral fermions beyond the SM ones. Indeed, a new multiplet of degenerate fermions will contribute to the value of the $S$ parameter in the following way~\cite{Amsler:2008zzb}:
\be
S = \frac{N_C}{3\pi} \sum_i \left( t_{3L}(i) - t_{3R}(i) \right)^2\,,
\ee
where $t_{3L}(i)$ and $t_{3R}(i)$ are the third components of weak isospins of the left-handed and the right-handed components of the fermion $i$, and $N_C$ is the number of colors. Considering an SM Higgs mass of $M_{H}$ = 117 GeV, the new physics contribution to the $S$ parameter is constrained to be $\lesssim 0.06$ at 95\% C.L.~\cite{Amsler:2008zzb}.

To be consistent with direct searches at collider experiments, we consider the mass of the triplet scalar mediator in the $s$-channel to be $\gtrsim 100$~GeV~\cite{Raspereza:2002ni}.

\subsubsection*{\noindent The annihilation cross section}
Using the Lagrangian terms of Eq.\,\eqref{eq:tripletNu} and Eq.\,\eqref{eq:trilinear} and the expression of the annihilation cross section given in Eq.\,\eqref{eq:S:scalar}, we find that
\bena
\sigma_{\rm ann} v=
 \frac{1}{8\pi}
\frac{\left(\gamma^{(2,3)}_{\chi \phi}\right)^2|Y^{\tau \tau}_{\nu;3}|^2}{(4 m_\chi^2-m_\phi^2)^2}
+\mathcal{O}(v^2) &\mbox{ for scalar DM, }
\label{eq:triplS}
\eena
where we have assumed for simplicity that the DM particle and the lightest neutral scalar mediator correspond to the real components of $\chi^0_{s;2}$ and $\phi^0_{s;3}$, respectively. The parameter $\gamma^{(2,3)}_{\chi \phi}$ is set to be real. Considering, instead, Eq.\,\eqref{eq:last} and Eq.\,\eqref{eq:F:scalar}, we conclude that
\bena
\sigma_{\rm ann} v=
 \frac{1}{4 \pi}
\frac{|Y_\chi|^2|Y^{\tau \tau}_{\nu;3}|^2}{(4 m_\chi^2-m_\phi^2)^2}\, m_\chi^2
+\mathcal{O}(v^2)  &\mbox{ for Majorana DM,}
\label{eq:triplF}
\eena
with
\be
Y_\chi=Y^{(2,2)}_{\chi;3}+\left[Y^{(2,2)}_{\chi;3}\right]^*\,.
\ee
We have assumed the Yukawa coupling $Y^{(2,2)}_{\chi;3}$ to dominate over $Y^{\prime(2,2)}_{\chi;3}$. If these two couplings are of the same order, instead, the first nonzero contribution to the annihilation cross section would be given by a $p$-wave term. Moreover, we have considered the imaginary component of the $\phi^0_{s;3}$ (which is a pseudoscalar!) as exchange particle. The real component would, as a true scalar, have a zero $s$-wave due to parity conservation.

The expressions reported above refer to the production of $\tau$-neutrinos. The DM annihilation into neutrinos with other flavours would be more suppressed, because of the bound reported in Eq.\,\eqref{eq:diagMM}, and can therefore be neglected.

\begin{figure}[t]
\begin{tabular}{cc}
\includegraphics[width=7cm,height=6cm]{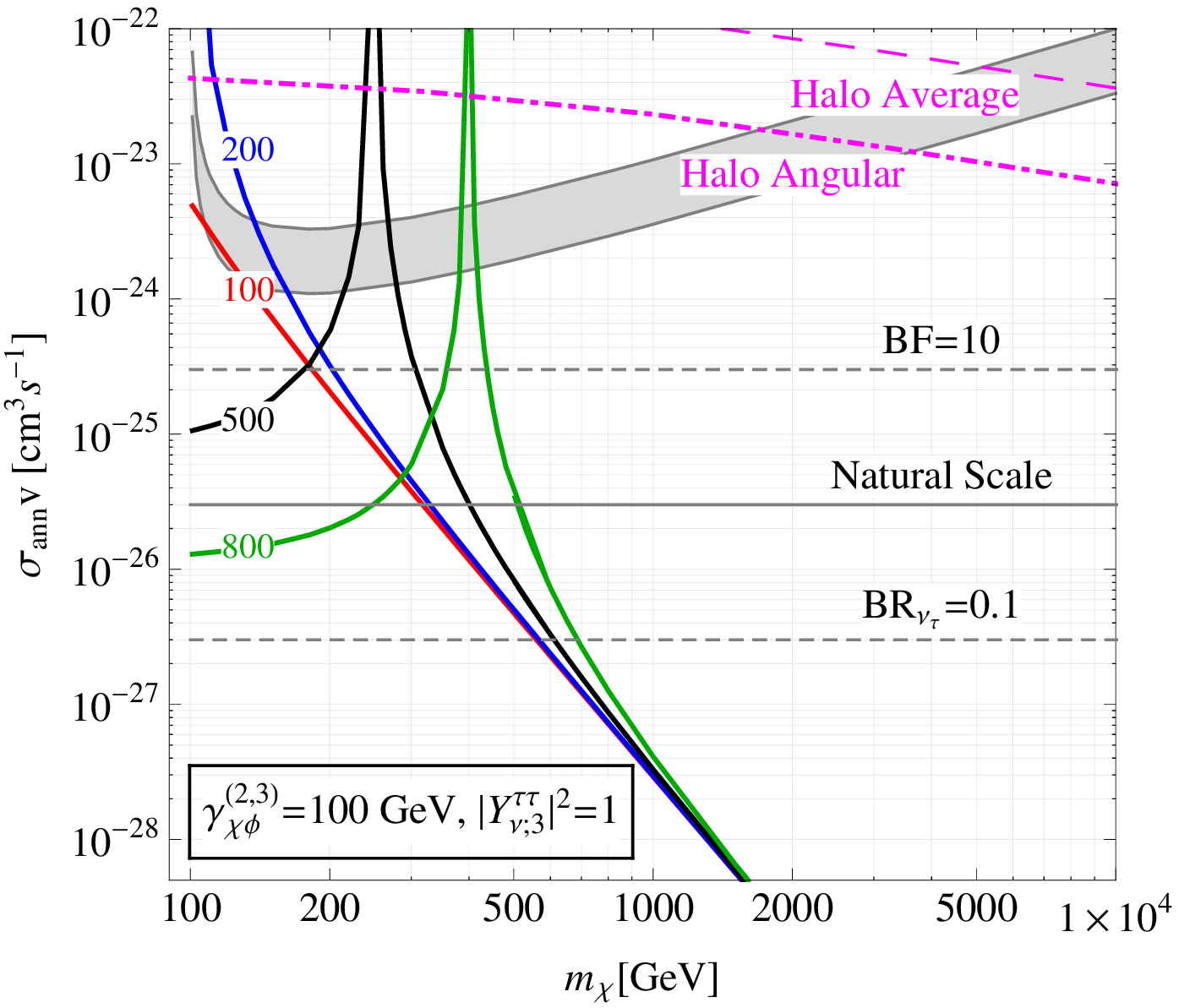}&
\includegraphics[width=7cm,height=6cm]{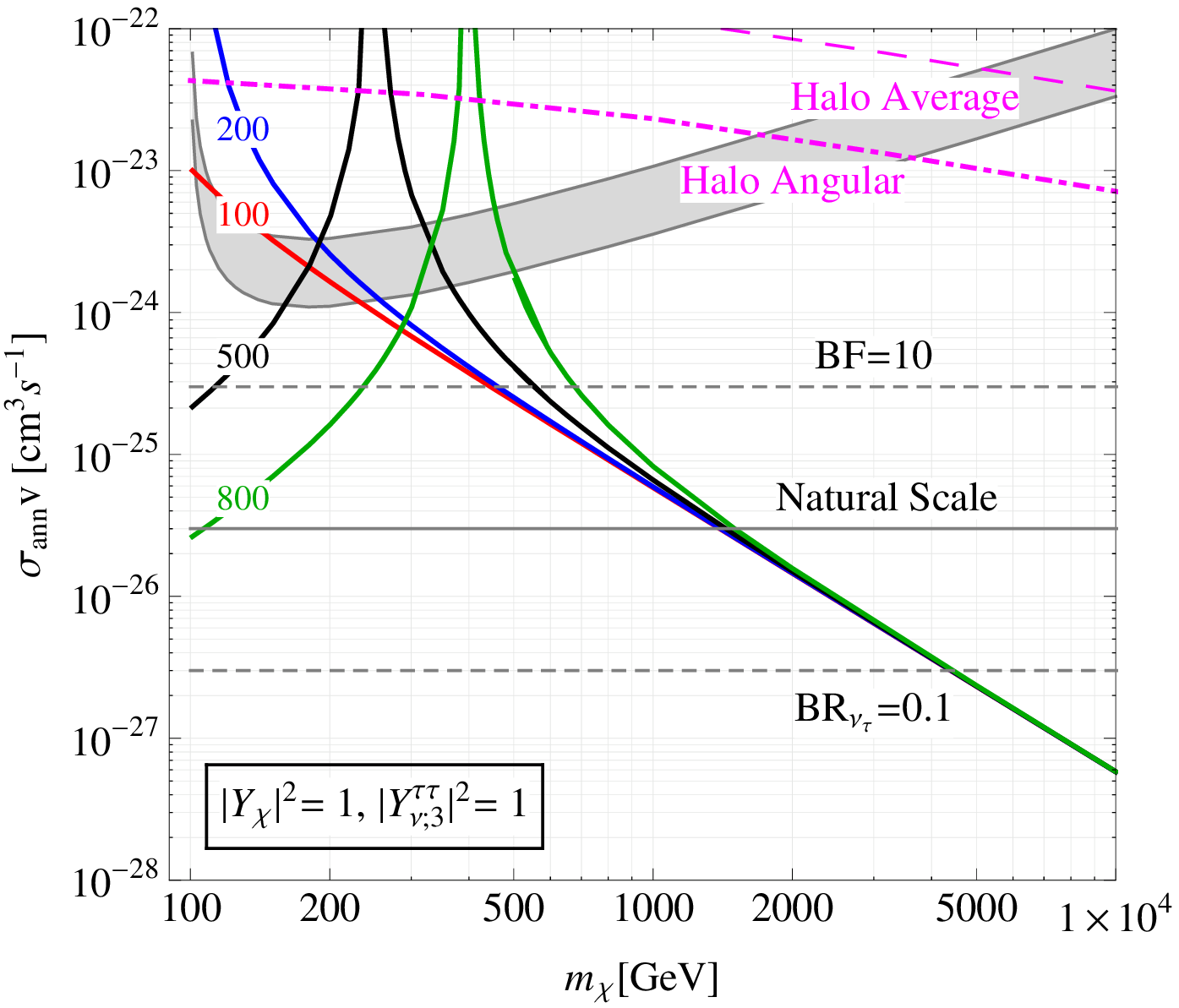}
\end{tabular}
\caption
{\label{fig:sigma_triplet} Dark Matter annihilation cross sections into tau neutrinos through the exchange of a scalar triplet with null VEV, in an $s$-channel diagram. Left panel: Scalar Dark Matter. Right panel: Majorana Dark Matter. The numbers next to each curve denote the different values of the scalar triplet mass (in GeV). The Halo Angular and the Halo Average lines represent bounds from neutrino searches, while the gray bands give the limits that could be set looking at the GC with a neutrino telescope located in the Northern Hemisphere, see text for more details. The horizontal gray solid line indicates the standard value of $\sigma_{\rm ann}v$ for a thermal relic (natural scale), while the gray dashed lines mark the values for a 10\% branching ratio into tau neutrinos (BR$_{\nu_\tau}$) and for a boost factor (BF) equal to ten (where the natural scale is taken as reference). }
\end{figure}

In Fig.\,\ref{fig:sigma_triplet}, we show the behavior of the annihilation cross section into tau neutrinos for the case of scalar DM (left panel) and of Majorana DM (right panel). The annihilation cross sections result to be of the order of the value expected for a thermal relic for a wide range of the parameter space. From our plots, it is possible to identify which are the values of the Yukawa couplings and of the triplet and DM mass in which the neutrino production might be relevant. This can then be applied to specific model, in which a triplet scalar without VEV is present.

The neutrino flux from the Galactic Center (GC), generated by the triplet scalar exchange, might be accessible to a future neutrino telescope located in the Northern Hemisphere only in the resonant region, in which $m_\chi \simeq m_\phi$. The gray bands in Fig.\,\ref{fig:sigma_triplet} indicate the limits (at 3$\sigma$ level) that could be set using contained muon events~\cite{Erkoca:2010qx} in a 1km$^3$ neutrino telescope with an energy threshold $E_\mu=100$ GeV. We have considered a cone half-angle of 30$^\circ$ around the GC and one year of exposure. The upper limit is obtained considering an isothermal DM density profile, while the lower one is derived from a NFW DM density profile. The limits that could be set using the IceCube detector are only of the order of $\sigma_{\rm ann} v \simeq 10^{-22}$~cm$^{-3}$~s$^{-1}$, as IceCube is not well-suited for looking at the Galactic Center, see Ref.\,\cite{Rott:2009hr} for more details. We finally wish to add that the CMB measurements of the WMAP satellite impose stringent limits on DM models with very large annihilation cross section, as has been pointed out in Refs.\,\cite{Galli:2009zc,Slatyer:2009yq}. Remember that even a DM particle that annihilates mainly into neutrinos will generally produce electromagnetic final states by loop diagrams~\cite{Dent:2008qy}. In considering specific DM models, the CMB recombination bounds of Refs.\,\cite{Galli:2009zc,Slatyer:2009yq} must be imposed.

The signals from the Sun and the Earth could, instead, be detected for a wide range of the parameters, depending on the value of the DM scattering cross section. For the Sun a $5\sigma$ discovery, after one year of data taking with the IceCube detector, can be achieved if $\sigma_p BR_{\nu}\simeq 6\cdot 10^{-7}$~pb for $m_\chi \simeq200$ GeV or if $\sigma_p BR_{\nu}\simeq 10^{-5}$~pb for $m_\chi \simeq 1$ TeV, where $\sigma_p$ is assumed to be dominated by spin-dependent interactions and where $BR_\nu$ is the branching ratio into neutrinos of all flavours~\cite{Barger:2007hj,Delaunay:2008pc}. For the Earth, assuming equilibrium between the capture and the annihilation rate, the $5\sigma$ discovery can be reached if $\sigma^{SI}_p BR_{\nu}\simeq 9\cdot 10^{-10}$~pb for $m_\chi \simeq200$~GeV and if $\sigma^{SI}_p BR_{\nu}\simeq 3\cdot 10^{-9}$~pb for $m_\chi \simeq 1$~TeV~\cite{Barger:2007hj,Delaunay:2008pc}

In the plots we also report the limits on the annihilation cross section $\sigma_{\rm ann} v$, as derived by the authors of Ref.\,\cite{Yuksel:2007ac} comparing the energy spectrum produced by DM pair-annihilation into neutrinos with the atmospheric neutrino background measured by the Super-Kamiokande, Frejus, and AMANDA detectors. The Halo Angular bound corresponds to a cone half-angle of about $30^\circ$ around the GC and to a value of the $J$-factor~\cite{Bergstrom:1997fj} of 25. The Halo Average bound is instead associated with $J\simeq5$, which is an average value for the whole sky. As can be seen from the figures, these constraints are not really strong and exclude only a small fraction of the parameter space in the resonance region.

Note that, for the case of Majorana DM that belongs to a doublet under $SU(2)_L$, the scalar triplet could also induce a neutrino production through a $t$-channel diagram. For simplicity, we show in Fig.\,\ref{fig:sigma_triplet} only the $s$-channel annihilation cross sections.

\subsubsection{$t$-channel: The singlet fermionic and scalar mediator}
\label{sec:Unsuppressed:singlet}

The couplings involved in a $t$-channel diagram are subject to experimental bounds, since they induce LFV processes at one loop. A summary of these experimental limits is given in the following, considering for definiteness the case of a singlet vector-like fermionic mediator $\phi_{f;1}$ (for a doublet scalar DM) and of a singlet scalar mediator $\phi_{s;1}$ (for a doublet vector-like DM). Bounds from measurements of the anomalous magnetic moments of the electron and the muon also apply.

\subsubsection*{\noindent Experimental constraints}
In the case of a scalar DM particle that is a doublet under $SU(2)_L$, the $\mu \to e \gamma$ process can be mediated by the charged scalar $\chi^{-}_{s;2}$ and the fermionic singlet $\phi_{f;1}$. 
Instead, for a doublet vector-like DM, the $\mu \to e \gamma$ process can be induced by the charged fermion $\chi^{-}_{f;2}$ and the scalar singlet $\phi_{s;1}$. Using the limit on $BR(\mu \to e \gamma)$ provided by the MEGA experiment~\cite{Ahmed:2001eh}, we can write
\be
3.2\cdot 10^9\, \frac{m^2_\mu/{\rm GeV}^2}{m^4_s/{\rm GeV}^4}\, \xi_1^4\, H^2(t)\,\lesssim 1.2 \cdot 10^{-11}\,,
\label{eq:MtoE}
\ee
where $m_\mu$ is the muon mass. We have defined $\xi_1^2 = \mathcal{T}^{(2,1)}_{e k} \left[\mathcal{T}^{(2,1)}_{k \mu}\right]^*$ and $t=m^2_f/m^2_s$, with $m_f$ and $m_s$ being, respectively, the mass of the fermion and the scalar particles involved in the loop process. The function $H(t)$ is given by~\cite{Lavoura:2003xp}
\be
H(t)=\left\{
 \begin{array}{rl}
 \frac{2 t^2+5 t -1}{12\,(t-1)^3}-\frac{t^2 \ln t}{2\,(t-1)^4} &\mbox{ for scalar DM,} \\
 \frac{t^2-5t-2}{12\,(t-1)^3}+\frac{t \ln t}{2\,(t-1)^4}  &\mbox{ for fermionic DM.}
 \end{array}
 \right.
\ee
In analogy, we find the following constraint on the couplings involved in the $\tau \to \mu \gamma$ process:
\be
2.1\cdot 10^6\, \frac{m^2_\tau/{\rm GeV}^2}{m^4_s/{\rm GeV}^4}\, \xi_2^4\, H^2(t)\,\lesssim 4.5 \cdot 10^{-8}\,,
\label{eq:TtoM}
\ee
where $m_\tau$ is the tau mass. We have defined $\xi_2^2 = \mathcal{T}^{(2,1)}_{e k} \left[\mathcal{T}^{(2,1)}_{k \tau}\right]^*$ and we have used the experimental limit on $BR(\tau \to \mu \gamma)$ as provided by the BELLE experiment~\cite{Raidal:2008jk}. Finally, the last bound coming from LFV is given by the BaBar~\cite{Aubert:2005wa} experimental limit on $BR(\tau \to e \gamma)$:
\be
2.1\cdot 10^6\, \frac{m^2_\tau/{\rm GeV}^2}{m^4_s/{\rm GeV}^4}\, \xi_3^4\, H^2(t)\,\lesssim 1.1 \cdot 10^{-7}\,,
\label{eq:TtoE}
\ee
where in this case we have $\xi_3^2 = \mathcal{T}^{(2,1)}_{\mu k} \left[\mathcal{T}^{(2,1)}_{k \tau}\right]^*$. Moreover the couplings $\mathcal{T}^{(2,1)}_{e k}$ and $\mathcal{T}^{(2,1)}_{\mu k}$ are also subject to constraints coming from measurements of the electron and the muon anomalous magnetic moments \cite{Fayet:2007ua}:
\bena
|\mathcal{T}^{(2,1)}_{e k}| \lesssim \mathcal{O}(10^{-4})\,\left(\frac{m_\phi}{{\rm MeV}}\right)\,,\\
|\mathcal{T}^{(2,1)}_{\mu k}| \lesssim \mathcal{O}(10^{-6})\,\left(\frac{m_\phi}{{\rm MeV}}\right)\,.
\eena
For simplicity, in our numerical examples we consider the situation in which $\mathcal{T}^{(2,1)}_{e k} \simeq \mathcal{T}^{(2,1)}_{\mu k} \ll \mathcal{T}^{(2,1)}_{\tau k}$. In this case, using Eq.\,\eqref{eq:MtoE} and Eq.\,\eqref{eq:TtoM}, we find the following constraint:
\be
|\mathcal{T}^{(2,1)}_{\tau k}|^2\lesssim
{\rm min}\left(1,\,8.7 \cdot 10^{-4}
\frac{m^2_s/{\rm GeV}^2}{m_\tau/{\rm GeV}}\, \frac{1}{H(t)}\right)\,,
\label{eq:tri}
\ee
where we have explicitly imposed that the coupling is at most of order one.

For the $t$-channel diagram, we restrict our analysis to a singlet Majorana mediator and to a singlet scalar mediator with masses $\gtrsim 100$ GeV. We use the limit of Eq.\,\eqref{eq:tri} in the numerical evaluation, considering also that in the case of scalar DM $m_s\simeq \mathcal{O}(m_\chi)$ and $m_f=m_\phi$, while in the case of fermionic DM $m_s=m_\phi$ and $m_f \simeq \mathcal{O}(m_\chi)$. We finally wish to add that, in the case of a chiral mediator $\phi_{f;1}$ or of a chiral DM $\chi_{f;2}$, the constraints from LFV processes might be much stronger. In particular, in Eq.\,\eqref{eq:tri}, we would have the mass of the fermionic particle exchanged in the loop instead of the tau mass. This can be related to a chirality flip in the fermionic line.

\subsubsection*{\noindent The annihilation cross section}
Using the Lagrangian term of Eq.\,\eqref{eq:tScalarDM} and the expression of the annihilation cross section given in Eq.\,\eqref{eq:S:maj}, we find that
\bena
\sigma_{\rm ann} v=
 \frac{1}{8\pi}
\frac{|\mathcal{T}^{(2,1)}_{\tau k}|^4}{(m_\chi^2+m_\phi^2)^2}\,m_\phi^2
+\mathcal{O}(v^2) &\mbox{ for scalar DM, }
\eena
where we have assumed that the fermionic mediator is a Majorana particle and that the DM is the real component of $\chi^0_{s;2}$. This can be, for example, the case of sneutrino annihilation through a neutralino exchange. Considering, instead, Eq.\,\eqref{eq:tFermionicDM} and Eq.\,\eqref{eq:F:maj}, we conclude that
\bena
\sigma_{\rm ann} v=
 \frac{1}{64\pi}
\frac{|\mathcal{T}^{(2,1)}_{\tau k}|^4}{(m_\chi^2+m_\phi^2)^2}\, m_\chi^2
+\mathcal{O}(v^2)  &\mbox{ for Majorana DM,}
\eena
where we have considered the real component of $\phi_{s;1}$ to be the lightest scalar mediator. Remember that, if the Majorana particle is the supersymmetric neutralino, the couplings $\mathcal{T}^{(2,1)}_{\tau k}$ will be proportional to the neutrino mass and thus the annihilation cross section into neutrinos will be negligible, see the discussion after Eq.\,\eqref{eq:neutralino}. In a more general model, however, the couplings are not fixed and the neutrino production can be sizable even if the DM particle is Majorana. This possibility is often overlooked in the literature.

The expressions reported above refer to the production of tau neutrinos, which we have assumed to be the dominant channel. Depending on the structure of the matrix $\mathcal{T}_{\alpha k}$, the other neutrino flavours could lead to sizable contributions. Nevertheless, the total annihilation cross section into neutrinos would be of the same order as the one obtained considering the tau neutrino as the dominant flavour channel.

\begin{figure}[t]
\begin{tabular}{cc}
\includegraphics[width=7cm,height=6cm]{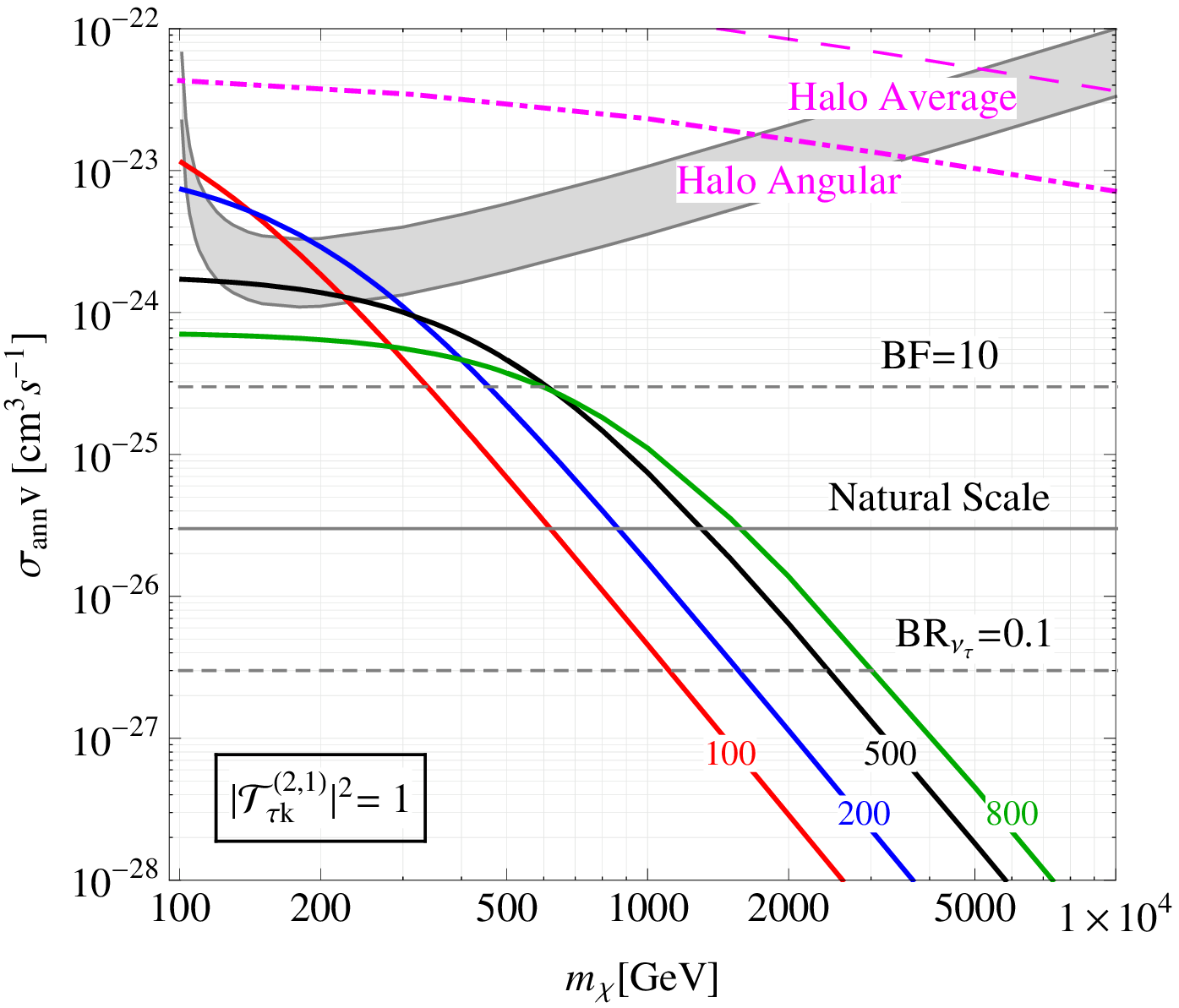}&
\includegraphics[width=7cm,height=6cm]{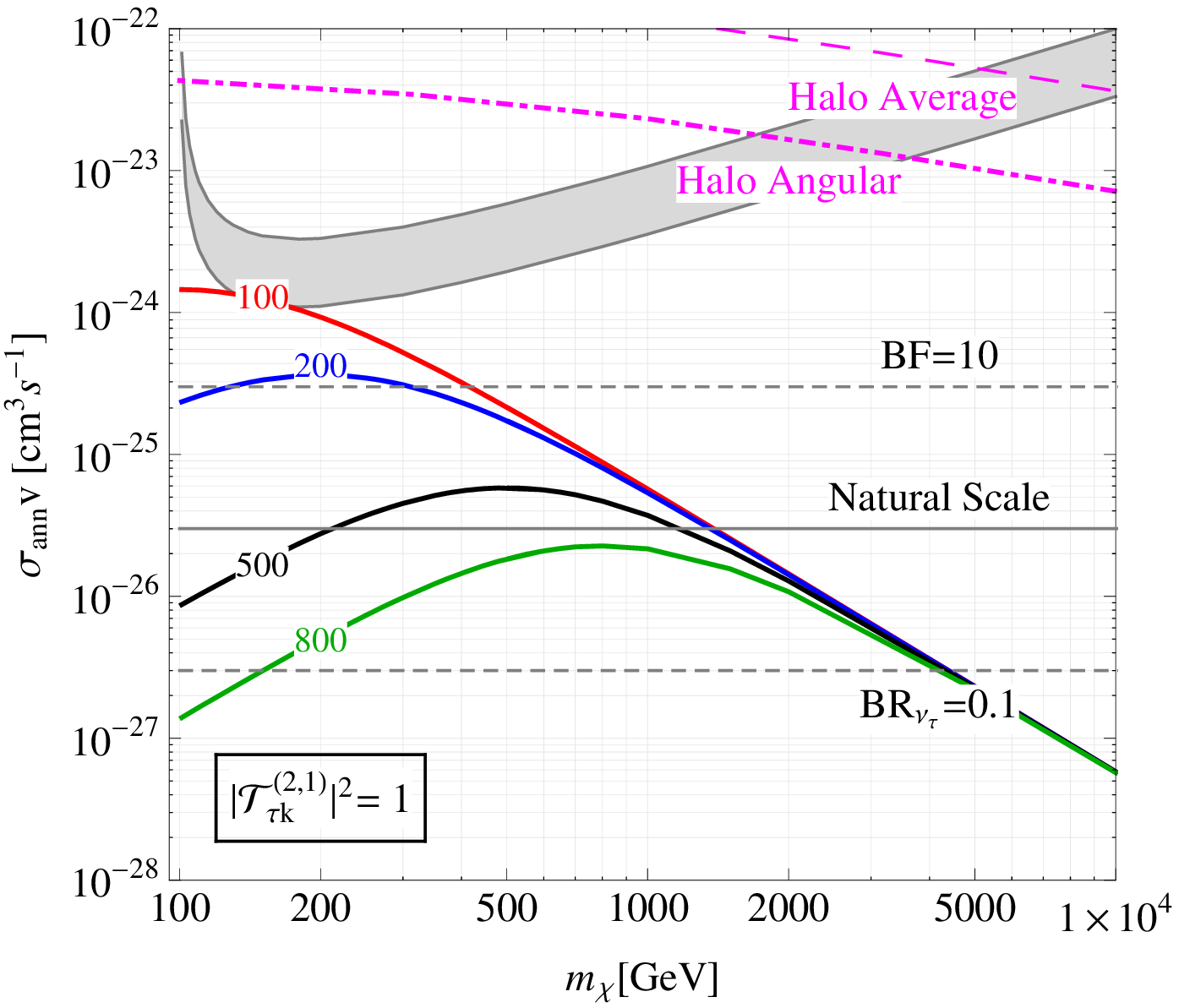}
\end{tabular}
\caption
{\label{fig:sigma_singlet} Dark Matter annihilation cross section into tau neutrinos through the exchange of a singlet mediator, in a $t$-channel diagram. Left panel: Scalar Dark Matter and Majorana mediator. Right panel: Majorana Dark Matter and scalar mediator. The numbers next to each curve denote the different values of the singlet mediator mass (in GeV). The Halo Angular and the Halo Average lines represent bounds from neutrino searches, while the gray bands give the limits that could be set looking at the GC with a neutrino telescope located in the Northern Hemisphere, see text for more details. The horizontal gray solid line indicates the standard value of $\sigma_{\rm ann}v$ for a thermal relic (natural scale), while the gray dashed lines mark the values for a 10\% branching ratio into tau neutrinos (BR$_{\nu_\tau}$) and for a boost factor (BF) equal to ten (where the natural scale is taken as reference).}
\end{figure}

The behavior of the annihilation cross sections into $\tau$-neutrinos is reported in Fig.\,\ref{fig:sigma_singlet} for the cases of scalar DM (left panel) and Majorana DM (right panel). For a wide range of the parameter space, the annihilation cross sections can cover the order of magnitudes expected for a standard WIMP. In our specific examples, the experimental limits on LFV processes reported in Eq.\,\eqref{eq:tri} result to be quite weak and do not restrict the allowed parameter space in the interesting region of $\sigma_{\rm ann} v$. However, we want to stress that in the case of a chiral mediator (for scalar DM) or in the case of a chiral fermionic DM, the bounds from LFV processes might be much stronger. In the plots we also report the Halo Angular and Halo Average bounds~\cite{Yuksel:2007ac}, which partially limit the regions of the annihilation cross section under consideration.

The neutrino signal from the GC, generated by a $t$-channel singlet exchange, could be hardly accessible to 
a future KM3Net-like neutrino telescope after one year of exposure, since a cross section of the order of $\gtrsim$~$10^{-24}$~cm$^3$~s$^{-1}$ is almost never reached. The signal from the Sun and the Earth, instead, might be detected, depending on the value of the scattering cross section, as we have explained in Sec.\,\ref{sec:Unsuppressed:triplet}.

\section{Conclusions}
\label{sec:conc}

We have performed a detailed model-independent analysis of the process of Dark Matter annihilation directly into neutrinos. Contrarily to some statements given in the literature, this channel is not always suppressed, depending on the neutrino mass models. In particular for scalar mediators, it might well be possible to find annihilation rates in a detectable region. Even though the neutrino mass is practically zero for all annihilation diagrams, the mechanism responsible for its generation can very well be decisive, since different mechanisms are constrained in different ways. We have shown how to systematically search for promising situations and how some cases relate to known models. Furthermore, we have illustrated two example considerations numerically. Further advantageous cases could be modelled in the future.

\section*{Acknowledgements}
\label{sec:ack}

We would like to thank A.~Adulpravitchai, E.~Akhmedov, W.~Rodejohann, and T.~Schwetz for useful discussions. This work has been supported by the DFG-Sonderforschungsbereich Transregio 27 `Neutrinos and beyond -- Weakly interacting particles in Physics, Astrophysics and Cosmology'.

\section*{Appendix A: Cross section formulae}
\label{app:ann}

\renewcommand{\theequation}{A-\arabic{equation}}
\setcounter{equation}{0}  

\pagestyle{headings}
The differential annihilation cross section of two DM particles $\chi$ into two neutrinos is given by~\cite{Dreiner:2008tw}
\be
v\,\frac{d\sigma_{\rm ann}}{d\cos \theta^*}=
\frac{1}{16\pi}\,\frac{1}{s}\,\overline{|\mathcal{M}|^{2}}\,,
\ee
where $v$ is the relative velocity between the two DM particles, $\theta^{*}$ is the scattering angle in the center-of-mass frame, $s \simeq 4 m^{2}_{\chi}+m^{2}_{\chi} v^{2}$ is the Mandelstam invariant, and $m_\chi$ is the DM mass. In the previous formula we have neglected the neutrino mass and we
have denoted the spin-averaged matrix element by $\overline{|\mathcal{M}|^{2}}$:
\be
\overline{|\mathcal{M}|^{2}}=\frac{1}{(2 S_{\rm DM}+1)^2}\sum_{\textrm{spins}}|\mathcal{M}|^{2}\,,
\ee
where $S_{\rm DM}$ is the spin of the DM particle and $\mathcal{M}$ is the total amplitude of the annihilation process under consideration.

Since we do not focus on a particular model, our results are general and can be applied to the calculation of the annihilation cross section into neutrinos for a specific DM candidate. Moreover, from our expressions it is easy to see which are the channels and the possible cases that could lead to a sizable DM branching ratio into neutrinos. For simplicity, throughout our analysis we consider only the Standard Model as gauge group. For the calculation we have used the {\tt FeynCalc} package~\cite{Mertig:1990an}.

\subsection*{Scalar Dark Matter}

For a scalar DM, the neutrino production can occur through a scalar and a $Z$-boson exchange in an $s$-channel diagram and through a fermion exchange in a $t$-channel diagram. In case the neutrinos are Majorana particles, also a $u$-channel diagram will be present.

\subsubsection*{\noindent Scalar mediator, $s$-channel}
Indicating the coupling of the scalar mediator to the DM with $D$ and the coupling to Dirac neutrinos with $N_{L} P_{L}+ N_{R} P_{R}$, with the projection operators defined as $P_{L,R}=(1\mp \gamma_{5})/2$, the total annihilation cross section can be written as
\bena
\sigma_{\rm ann} v\,(\chi_s;\phi_s;s)=
\frac{2^{-n}}{8 \pi}|D|^{2}
\frac{|N_{L}|^{2}+|N_{R}|^{2}}{(4 m^{2}_{\chi}-m^{2}_{\phi})^{2}}\,4^n
\left(1-\frac{2 m^{2}_{\chi}}
{(4m^{2}_{\chi}-m^{2}_{\phi})}v^{2}
\right)+\mathcal{O}(v^{4})\,,
\label{eq:S:scalar}
\eena
where $m_\phi$ is the scalar mediator mass and $n=0,1$ for Dirac and Majorana neutrinos, respectively.
For Majorana neutrinos $\nu^M$, anti-symmetrization of the final states must be imposed. In that case, a factor 1/2 will be present to avoid double counting of identical particles in the final state, and a factor 4 arises from the Feynman rule for the effective vertex since the Majorana neutrinos are self-conjugate particles. Note that, for simplicity, we use the same form of the Yukawa couplings for Dirac and Majorana neutrinos. We want to stress that in general this is not the case, since Dirac and Majorana neutrinos usually couple to scalar mediators with different $SU(2)$ representations, see Secs.~\ref{sec:nuprod} and~\ref{sec:summary}.

\subsubsection*{\noindent $Z$-boson mediator, $s$-channel}
Indicating the coupling of the $Z$-boson to the DM generically as $D (k_{1}-k_{2})^{\mu}$, with $k_{1}$ and $k_{2}$ being the DM 4-momenta, and the coupling to the neutrinos with $N_{L} \gamma^{\mu} P_{L}$, the total annihilation cross section can be written as
\be
\sigma_{\rm ann} v\,(\chi_s;Z;s)=\frac{1}{12 \pi}D^{2}\frac{N_{L}^{2}}
{(4 m^{2}_{\chi}-m^{2}_{Z})^{2}}\,m^{2}_{\chi}v^{2} + \mathcal{O}(v^{4})\,,
\label{eq:S:Z}
\ee
with $D$ and $N_L$ being real numbers. In this case the annihilation cross section is proportional to the DM velocity, as we would naively expect from angular momentum conservation. If the neutrinos are Majorana particles, the annihilation cross section is equivalent to the one given in Eq.\,\eqref{eq:S:Z}. Indeed, it is well known that weak interactions mediated by the $Z$-boson do not distinguish between Dirac and Majorana neutrinos~\cite{Kayser:1982br}.

\subsubsection*{\noindent Fermionic mediator, $t\&u$-channels}
Indicating the coupling of the DM particle to the fermionic mediator and the neutrino with $F_{L} P_{L} + F_{R} P_{R}$ at one vertex, and with $G_{L} P_{L} + G_{R} P_{R}$ at the other vertex, the total annihilation cross section is given by
\bena
\sigma_{\rm ann} v\,(\chi_s;\phi_f;t)&=&\frac{1}{8 \pi}
\frac{|F_{L}|^{2}|G_{L}|^{2}+|F_{R}|^{2}|G_{R}|^{2}}
{(m^{2}_{\chi}+m^{2}_{\phi})^{2}}
\left(m^{2}_{\phi}-
\frac{m_\phi^4}
{\left(m_\chi^2+m_\phi^2\right)^2} \,m_\chi^2 v^2 \right)+\nonumber\\
&+&\frac{1}{48 \pi}
\frac{|F_{R}|^{2} |G_{L}|^{2}+|F_{L}|^{2} |G_{R}|^{2}}
{\left(m_\chi^2+m_\phi^2\right)^2}\,
 m_\chi^2 v^2+O\left(v^4\right)\,,
\label{eq:S:t}
\eena
where $m_\phi$ is the fermionic mediator mass. Notice that, in general, $G_L= F_R^*$ and $G_R=F_L^*$ for a Dirac mediator, while also $G_L= F_L$ and $G_R=F_R$ are allowed for a Majorana mediator. In the first case a pair of $\nu \bar{\nu}$ is produced, while in the second case $\nu \nu$ (or $\bar{\nu} \bar{\nu}$) are produced. If the DM particle is a real scalar, also a $u$ channel will be present. The corresponding cross section is equivalent to the one in Eq.\,\eqref{eq:S:t}.

In the case of Majorana neutrinos both, the $t$-channel and the $u$-channel diagram, must be considered and `added' with a relative minus sign. The annihilation cross section is then modified to
\bena
\sigma_{\rm ann} v\,(\chi_s;\phi_f;t\&u)&=&
\frac{1}{4 \pi}
\frac{|F_{L}|^{2} |G_{L}|^{2}+|F_{R}|^{2} |G_{R}|^{2}}
{\left(m_\chi^2 +m_\phi^2\right)^2}
\left( m_\phi^2 -\frac{m_\phi^2\,\left( 3 m_\phi^2+m_\chi^2\right)}
{3\left(m_\chi^2+m_\phi^2\right)^2}\,m_\chi^2 v^2 \right)+\nonumber\\
&+&\frac{1}{48 \pi}
\frac{|F_{R} G_{L}-F_{L} G_{R}|^2}
{\left(m_\chi^2+m_\phi^2\right)^2}\,m_\chi^2 v^2+O\left(v^4\right)\,.
\label{eq:S:maj}
\eena

\subsection*{Fermionic Dark Matter}

For fermionic DM, the neutrino production can occur through a $Z$-boson exchange in an $s$-channel diagram and through a scalar exchange in an $s$-channel or a $t$-channel diagram. In case the DM or the neutrinos are Majorana particles, also a $u$-channel diagram will be present.

\subsubsection*{\noindent Scalar mediator, $s$-channel}
Indicating the coupling of the Dirac DM particle to the scalar mediator with $D_{L} P_{L} + D_{R} P_{R}$ and the one of the Dirac neutrinos with $N_{L} P_{L} + N_{R} P_{R}$, the total annihilation cross section can be written as
\bena
\sigma_{\rm ann} v\,(\chi_f;\phi_s;s)&=&\frac{2^{-n}}{16 \pi}
\frac{|N_L|^2+ |N_R|^2}{(4 m_\chi^2 - m_\phi^2)^2 }\, 4^{n} 4^{m}\times \nonumber\\
&\times& \left( |D_L-D_R|^2\, m_\chi^2  -
\frac{m_\phi^2}{2\,(4 m_\chi^2 - m_\phi^2)}
(|D_L|^2+|D_R|^2)\,m_\chi^2  v^2 \right.+\nonumber\\
& & +\left.\frac{m_\chi^2}{8\,(4 m_\chi^2 - m_\phi^2)}
(D_L D_R^*+\,c.c.)\,m_\chi^2  v^2\right) +\mathcal{O} (v^4)\,,
\label{eq:F:scalar}
\eena
where $m_\phi$ is the scalar mediator mass, $n$=0 ($n$=1) for Dirac (Majorana) neutrinos and $m$=0 ($m$=1) for Dirac (Majorana) DM. The factor 1/2 is present to avoid double counting of identical particles in the final state, while the factors 4 come from the Feynman rules for the effective vertex. For simplicity, we have used the same Yukawa couplings for Dirac and Majorana neutrinos. However, they generally couple to different scalar particles, see Secs.~\ref{sec:nuprod} and~\ref{sec:summary}.

Note, that in case the DM couples to the scalar mediator through a scalar coupling (i.e.\ $D_L=D_R$), the cross section will be proportional to the DM velocity $v$. This is a consequence of parity conservation: A fermion-antifermion pair has a parity of ($-1$) and can therefore, in an $s$-wave configuration, only couple to a pseudoscalar particle (i.e.\ $D_L=-D_R$).

\subsubsection*{\noindent $Z$-boson mediator, $s$-channel}
Indicating the coupling of the DM particle to the $Z$-boson with $\gamma^{\mu} (D_{L} P_{L} + D_{R} P_{R}) $ and the one of the neutrino with $N_{L} \gamma^{\mu} P_{L} $, the total annihilation cross section can be written as
\bena
\sigma_{\rm ann} v\,(\chi^D_f;Z;s)&=& \frac{1}{8 \pi}
\frac{N_L^2 }{(4 m_\chi^2-m_Z^2)^2}\times \nonumber\\
&\times&\left((D_L+ D_R)^2 \, m_\chi^2 -
\frac{(m_Z^2+2 m_\chi^2)}{3\,(4 m_\chi^2-m_Z^2)}
(D_L^2+D_R^2) \, m_\chi^2  v^2 \right.-\nonumber\\
& &-\left.\frac{4\,m_\chi^2}{(4 m_\chi^2-m_Z^2)}
( D_L D_R )\,m_\chi^2  v^2\right)+
\mathcal{O}(v^4)\,,
\label{eq:F:Z}
\eena
with $D_L$, $D_R$, and $N_L$ being real numbers. The cross section for Majorana neutrinos is equivalent to Eq.\,\eqref{eq:F:Z}, since, as we have mentioned before, the weak interactions mediated by the $Z$-boson do not distinguish between Dirac and Majorana neutrinos~\cite{Kayser:1982br}.

If the DM particle is a Majorana fermion, the cross section reported above will be drastically modified. Indeed, in an $s$-wave annihilation, the fermions in the initial state are forced to have opposite spins by the Pauli exclusion principle. As a consequence, since the $Z$-boson has a spin of one, we expect that the first nonzero contribution to the annihilation cross section for Majorana DM will be given by the $p$-wave term. Indeed, we find:
\bena
\sigma_{\rm ann} v\,(\chi^M_f;Z;s) = \frac{1}{12 \pi }
\frac{\,N_L^2}{(4\,m_\chi^2 -m_Z^2)^2}\, (D_L -D_R)^2\,
m_\chi^2 v^2
+ \mathcal{O}\left(v^4\right)\,,
\eena
where the same expression holds in case the DM and the neutrinos are both Majorana particles.

\subsubsection*{\noindent Scalar mediator, $t\&u$-channels}
Indicating the coupling of the DM particle to the fermionic mediator and to the neutrino at one vertex with $F_{L} P_{L} + F_{R} P_{R}$ and at the other one with $G_{L} P_{L} + G_{R} P_{R}$, the total annihilation cross section is given by
\bena
\sigma_{\rm ann} v \,(\chi_f;\phi_s, t)&=& \frac{1}{32 \pi}
\frac{(|F_L|^2+|F_R|^2) (|G_L|^2+|G_R|^2)}{(m_\chi^2+m_\phi^2)^2}\times \nonumber\\
&\times&\left( m_\chi^2+
\frac{(m_\phi^4-3 m_\chi^2 m_\phi^2-m_\chi^4)}{3\,(m_\chi^2+m_\phi^2)^2}\,
m_\chi^2 v^2\right)+\mathcal{O}(v^4)\,,
\label{eq:F:t}
\eena
where $m_\phi$ is the fermionic mediator mass. Note that, in general, $G_L= F_R^*$ and $G_R=F_L^*$ or 
$G_L= F_L$ and $G_R=F_R$. In the first case a pair of $\nu \bar{\nu}$ is produced, while in the second case $\nu \nu$ (or $\bar{\nu} \bar{\nu}$) are produced.

A $t$-channel and a $u$-channel diagram must be considered in the case of Majorana neutrinos and/or Majorana DM. The expression for the annihilation cross section will thus be modified to
\bena
\sigma_{\rm ann} v\,(\chi_f;\phi_s;t\&u) &=&
\frac{1}{32 \pi}
\frac{1}{(m_\chi^2+m_\phi^2)^2}
\,m_\chi^2 \times \mathcal{A}-\nonumber \\
&-&\frac{1}{192 \pi}
\frac{1}{(m_\chi^2+m_\phi^2)^4} m_\chi^2 v^2 \times \mathcal{B}+ \mathcal{O}\left(v^4\right)\,.
\label{eq:F:maj}
\eena
The functions $\mathcal{A}$ and $\mathcal{B}$ are given by the following expressions in the case of Majorana neutrinos:
\bena
\mathcal{A_\nu}&=&
|F_L|^2|G_L|^2+|F_R|^2|G_R|^2 + |F_L G_R-F_R G_L|^2\,,
\eena
\bena
\mathcal{B_\nu}&=& (|F_L|^2 |G_L|^2 + |F_R|^2|G_R|^2)\,
( m_\chi^4 + 4 m_\chi^2 m_\phi^2 - 3 m_\phi^4 )+\nonumber \\
&+&(|F_L|^2 |G_R|^2 + |F_R|^2|G_L|^2)\,
( m_\phi^4 - 3 m_\chi^2 m_\phi^2 -  m_\chi^4 )-\nonumber \\
&-& 2\, (F_L F_R^* G_L^*G_R+\,c.c.)\,(3 m_\phi^2 + 2 m_\chi^2)\,m_\chi^2\,.
\eena
In the case of Majorana DM, the corresponding expressions for $\mathcal{A}$ and $\mathcal{B}$ are given by
\bena
\mathcal{A_\chi}&=&
2\,|F_L|^2|G_L|^2+2\,|F_R|^2|G_R|^2\,,
\label{eq:neutralino}
\eena
\bena
\mathcal{B_\chi}&=& 2\,(|F_L|^2 |G_L|^2 + |F_R|^2|G_R|^2)\,
(3 m_\phi^4 - 4 m_\chi^2 m_\phi^2 - m_\chi^4 )+\nonumber \\
&+&4\,(|F_L|^2 |G_R|^2 + |F_R|^2|G_L|^2)\,
(m_\phi^4+ m_\chi^4 )\,.
\eena
Notice that, for Majorana DM, terms proportional to $F_L G_R$ or $F_R G_L$ are not present in the $s$-wave. Indeed, due to the Pauli principle, two Majorana particles cannot have parallel spins if their relative angular momentum $l$ is zero. The only nonzero contribution to the $s$-wave configuration will be present if $F_L \ne 0$ and $G_L \ne 0$. This situation can arise in supersymmetric models only in the presence of a mixing between the left and right sfermions. However, a mixing term between $\tilde{f}_L$ and $\tilde{f}_R$ is proportional to the fermion mass. For this reason, the annihilation cross section of a neutralino pair into fermions through a $t$-channel sfermion exchange is always proportional to the mass of the fermions produced. This conclusion does, however, not hold in general when we consider a Majorana DM beyond a supersymmetric framework.

\subsection*{Vector Dark Matter}

In this Section, we report the annihilation cross sections for the case of vector DM, since in specific models, for example in theories with Extra Dimensions, new vector particles can be present even without extending the SM gauge group. The neutrino production can then occur through a scalar exchange in an $s$-channel diagram and through a fermion exchange in a $t$-channel diagram. In case the neutrinos are Majorana particles, also a $u$-channel diagram will be present. However, the coupling in the case of a $Z$-boson mediator can only exist in theories with an extended gauge group.

\subsubsection*{\noindent Scalar mediator, $s$-channel}
Indicating the coupling of the scalar mediator to the DM particles with $D$ and the one to the neutrinos with $N_{L} P_{L} + N_{R} P_{R}$, the total annihilation cross section can be written as
\bena
\sigma_{\rm ann} v \,(\chi_v;\phi_s;s)=\frac{2^{-n}}{24 \pi}D^{2}
\frac{N_L^2+N_R^{2}}{(4 m^{2}_{\chi}-m^{2}_{\phi})^{2}}\,4^n
\left(1-\frac{(2 m^{2}_{\chi}+ m^{2}_{\phi})}
{3\,(4m^{2}_{\chi}-m^{2}_{\phi})}v^{2}
\right)+\mathcal{O}(v^{4})\,,
\label{eq:V:Scalar}
\eena
with $D$, $N_L$, and $N_R$ being real numbers, and $n=0,1$ for Dirac and Majorana neutrinos, respectively. As in the previous cases, we have considered the same couplings for Dirac and Majorana neutrinos for simplicity.

\subsubsection*{\noindent $Z$-boson mediator, $s$-channel}
Indicating the coupling of the $Z$-boson to the DM generically as $D (k_{1}-k_{2})^{\mu}$, with $k_{1}$ and $k_{2}$ being the DM 4-momenta, and the coupling to the neutrinos with $N_{L} \gamma^{\mu} P_{L}$, the total annihilation cross section can be written as
\bena
\sigma_{\rm ann} v \,(\chi_v;Z;s)=\frac{1}{9 \pi}D^{2}\frac{N_{L}^{2}}
{(4 m^{2}_{\chi}-m^{2}_{Z})^{2}}\,m^{2}_{\chi}v^{2}+\mathcal{O}(v^{4})\,,
\label{eq:V:Vector}
\eena
with $D$ and $N_L$ being real numbers. As to be expected, the derivative coupling of the 3-vector vertex results into proportionality to the DM velocity.

\subsubsection*{\noindent Fermionic mediator, $t\&u$-channels}
Indicating the coupling of the DM particle to the fermionic mediator and to the neutrino at one vertex with $\gamma^\mu(F_{L} P_{L} + F_{R} P_{R})$ and at the other one with $\gamma^\nu(G_{L} P_{L} + G_{R} P_{R})$, the total annihilation cross section is given by
\bena
\sigma_{\rm ann} v \,(\chi_v;\phi_f;t)&=& \frac{1}{72 \pi}
\frac{4 m_\chi^2 (F_L^2 G_L^2+F_R^2 G_R^2) +
5 m_\phi^2 (F_R^2 G_L^2+F_L^2 G_R^2) }{(m_\chi^2+m_\phi^2)^2} +\nonumber \\
&+&\frac{1}{432 \pi}
\frac{1}{(m_\chi^2+m_\phi^2)^4} \,v^2 \times \mathcal{C}+ \mathcal{O}\left(v^4\right)\,,
\label{eq:V:t}
\eena
with
\bena
\mathcal{C}&=& 12\, m_\phi^6\, (F_L^2 G_R^2 + F_R^2 G_L^2) +
13\, m_\chi^6\, (F_L^2 G_L^2 + F_R^2 G_R^2) + \nonumber \\
&+& m_\chi^2\, m_\phi^4 \, (13 F_L^2 G_L^2 + 13 F_R^2 G_R^2
+ 2 F_L^2 G_R^2 + 2 F_R^2 G_L^2) + \nonumber \\
&+&2\,m_\chi^4\, m_\phi^2 \,(F_L^2 G_L^2 + F_R^2 G_R^2) +
20\,m_\chi^4\, m_\phi^2 \,(F_L^2 G_R^2 + F_R^2 G_L^2)\,.
\eena
Note that, in general, $G_L= F_L$ and $G_R=F_R$. For Majorana neutrinos, a $t$-channel and a $u$-channel diagram are present. The annihilation cross section is then modified to
\bena
\sigma_{\rm ann} v\,(\chi_v;\phi_f; t\&u)&=& \frac{1}{36 \pi}
\frac{ (G_L^2 + G_R^2) (2 m_\chi^2 F_L^2 + 3 m_\phi^2 F_R^2)+4 m_\chi^2\, G_L G_R F_L F_R }
{(m_\chi^2+m_\phi^2)^2}+\nonumber \\
&+&\frac{1}{432  \pi}
\frac{1}{(m_\chi^2+m_\phi^2)^4} \,v^2 \times \mathcal{D}+ \mathcal{O}\left(v^4\right)\,,
\label{eq:V:maj}
\eena
with
\bena
\mathcal{D}&=& 12\, m_\phi^6\, (F_L^2 G_R^2 + F_R^2 G_L^2) +
13\, m_\chi^6\, (F_L^2 G_L^2 + F_R^2 G_R^2) + \nonumber \\
&+& 13\,m_\chi^2\, m_\phi^4 \, ( F_L^2 G_L^2  + F_R^2 G_R^2) -
4\, m_\chi^2 \, m_\phi^4 \, ( F_L^2 G_R^2 + F_R^2 G_L^2) + \nonumber \\
&+&2\,m_\chi^4\, m_\phi^2 \,( F_L^2 G_L^2 + F_R^2 G_R^2) +
16\,m_\chi^4\, m_\phi^2 \, ( F_L^2 G_R^2 + F_R^2 G_L^2) - \nonumber \\
&-& 2 \, F_L F_R G_L G_R \,(9 m_\chi^4 + 10 m_\chi^2 m_\phi^2 - 7 m_\phi^4)\,.
\eena

\bibliographystyle{./apsrev}
\bibliography{./DM-Ann}

\begin{thebibliography}{10}
\expandafter\ifx\csname bibnamefont\endcsname\relax
  \def\bibnamefont#1{#1}\fi
\expandafter\ifx\csname bibfnamefont\endcsname\relax
  \def\bibfnamefont#1{#1}\fi
\expandafter\ifx\csname url\endcsname\relax
  \def\url#1{\texttt{#1}}\fi
\expandafter\ifx\csname urlprefix\endcsname\relax\def\urlprefix{URL }\fi
\providecommand{\bibinfo}[2]{#2}
\providecommand{\eprint}[2][]{\url{#2}}

\bibitem{Hooper:2008cf}
\bibinfo{author}{\bibfnamefont{D.}~\bibnamefont{Hooper}},
  \bibinfo{author}{\bibfnamefont{F.}~\bibnamefont{Petriello}},
  \bibinfo{author}{\bibfnamefont{K.~M.} \bibnamefont{Zurek}}, \bibnamefont{and}
  \bibinfo{author}{\bibfnamefont{M.}~\bibnamefont{Kamionkowski}},
  \bibinfo{journal}{Phys. Rev.} \textbf{\bibinfo{volume}{D79}},
  \bibinfo{pages}{015010} (\bibinfo{year}{2009}), \eprint{0808.2464}.

\bibitem{Feng:2008qn}
\bibinfo{author}{\bibfnamefont{J.~L.} \bibnamefont{Feng}},
  \bibinfo{author}{\bibfnamefont{J.}~\bibnamefont{Kumar}},
  \bibinfo{author}{\bibfnamefont{J.}~\bibnamefont{Learned}}, \bibnamefont{and}
  \bibinfo{author}{\bibfnamefont{L.~E.} \bibnamefont{Strigari}}
  (\bibinfo{year}{2008}), \eprint{0808.4151}.

\bibitem{Kumar:2009ws}
\bibinfo{author}{\bibfnamefont{J.}~\bibnamefont{Kumar}},
  \bibinfo{author}{\bibfnamefont{J.~G.} \bibnamefont{Learned}},
  \bibnamefont{and} \bibinfo{author}{\bibfnamefont{S.}~\bibnamefont{Smith}}
  (\bibinfo{year}{2009}), \eprint{0908.1768}.

\bibitem{Niro:2009mw}
\bibinfo{author}{\bibfnamefont{V.}~\bibnamefont{Niro}},
  \bibinfo{author}{\bibfnamefont{A.}~\bibnamefont{Bottino}},
  \bibinfo{author}{\bibfnamefont{N.}~\bibnamefont{Fornengo}}, \bibnamefont{and}
  \bibinfo{author}{\bibfnamefont{S.}~\bibnamefont{Scopel}},
  \bibinfo{journal}{Phys. Rev.} \textbf{\bibinfo{volume}{D80}},
  \bibinfo{pages}{095019} (\bibinfo{year}{2009}), \eprint{0909.2348}.

\bibitem{Goldberg:1983nd}
\bibinfo{author}{\bibfnamefont{H.}~\bibnamefont{Goldberg}},
  \bibinfo{journal}{Phys. Rev. Lett.} \textbf{\bibinfo{volume}{50}},
  \bibinfo{pages}{1419} (\bibinfo{year}{1983}).

\bibitem{Jungman:1995df}
\bibinfo{author}{\bibfnamefont{G.}~\bibnamefont{Jungman}},
  \bibinfo{author}{\bibfnamefont{M.}~\bibnamefont{Kamionkowski}},
  \bibnamefont{and} \bibinfo{author}{\bibfnamefont{K.}~\bibnamefont{Griest}},
  \bibinfo{journal}{Phys. Rept.} \textbf{\bibinfo{volume}{267}},
  \bibinfo{pages}{195} (\bibinfo{year}{1996}), \eprint{hep-ph/9506380}.

\bibitem{Kakizaki:2003jk}
\bibinfo{author}{\bibfnamefont{M.}~\bibnamefont{Kakizaki}},
  \bibinfo{author}{\bibfnamefont{Y.}~\bibnamefont{Ogura}}, \bibnamefont{and}
  \bibinfo{author}{\bibfnamefont{F.}~\bibnamefont{Shima}},
  \bibinfo{journal}{Phys. Lett.} \textbf{\bibinfo{volume}{B566}},
  \bibinfo{pages}{210} (\bibinfo{year}{2003}), \eprint{hep-ph/0304254}.

\bibitem{Petcov:2009zr}
\bibinfo{author}{\bibfnamefont{S.~T.} \bibnamefont{Petcov}},
  \bibinfo{author}{\bibfnamefont{H.}~\bibnamefont{Sugiyama}}, \bibnamefont{and}
  \bibinfo{author}{\bibfnamefont{Y.}~\bibnamefont{Takanishi}},
  \bibinfo{journal}{Phys. Rev.} \textbf{\bibinfo{volume}{D80}},
  \bibinfo{pages}{015005} (\bibinfo{year}{2009}), \eprint{0904.0759}.

\bibitem{Agrawal:2010fh}
\bibinfo{author}{\bibfnamefont{P.}~\bibnamefont{Agrawal}},
  \bibinfo{author}{\bibfnamefont{Z.}~\bibnamefont{Chacko}},
  \bibinfo{author}{\bibfnamefont{C.}~\bibnamefont{Kilic}}, \bibnamefont{and}
  \bibinfo{author}{\bibfnamefont{R.~K.} \bibnamefont{Mishra}}
  (\bibinfo{year}{2010}), \eprint{1003.1912}.

\bibitem{Falkowski:2009yz}
\bibinfo{author}{\bibfnamefont{A.}~\bibnamefont{Falkowski}},
  \bibinfo{author}{\bibfnamefont{J.}~\bibnamefont{Juknevich}},
  \bibnamefont{and} \bibinfo{author}{\bibfnamefont{J.}~\bibnamefont{Shelton}}
  (\bibinfo{year}{2009}), \eprint{0908.1790}.

\bibitem{Blennow:2009ag}
\bibinfo{author}{\bibfnamefont{M.}~\bibnamefont{Blennow}},
  \bibinfo{author}{\bibfnamefont{H.}~\bibnamefont{Melbeus}}, \bibnamefont{and}
  \bibinfo{author}{\bibfnamefont{T.}~\bibnamefont{Ohlsson}}
  (\bibinfo{year}{2009}), \eprint{0910.1588}.

\bibitem{Akhmedov:1999uz}
\bibinfo{author}{\bibfnamefont{E.~K.} \bibnamefont{Akhmedov}}
  (\bibinfo{year}{1999}), \eprint{hep-ph/0001264}.

\bibitem{Grimus:2003es}
\bibinfo{author}{\bibfnamefont{W.}~\bibnamefont{Grimus}},
  \bibinfo{journal}{Lect. Notes Phys.} \textbf{\bibinfo{volume}{629}},
  \bibinfo{pages}{169} (\bibinfo{year}{2004}), \eprint{hep-ph/0307149}.

\bibitem{Barger:2003qi}
\bibinfo{author}{\bibfnamefont{V.}~\bibnamefont{Barger}},
  \bibinfo{author}{\bibfnamefont{D.}~\bibnamefont{Marfatia}}, \bibnamefont{and}
  \bibinfo{author}{\bibfnamefont{K.}~\bibnamefont{Whisnant}},
  \bibinfo{journal}{Int. J. Mod. Phys.} \textbf{\bibinfo{volume}{E12}},
  \bibinfo{pages}{569} (\bibinfo{year}{2003}), \eprint{hep-ph/0308123}.

\bibitem{Mohapatra:2005wg}
\bibinfo{author}{\bibfnamefont{R.~N.} \bibnamefont{Mohapatra}} \emph{et~al.},
  \bibinfo{journal}{Rept. Prog. Phys.} \textbf{\bibinfo{volume}{70}},
  \bibinfo{pages}{1757} (\bibinfo{year}{2007}), \eprint{hep-ph/0510213}.

\bibitem{Giunti:2007ry}
\bibinfo{author}{\bibfnamefont{C.}~\bibnamefont{Giunti}} \bibnamefont{and}
  \bibinfo{author}{\bibfnamefont{C.~W.} \bibnamefont{Kim}},
  \emph{\bibinfo{title}{{Fundamentals of Neutrino Physics and Astrophysics}}},
  \bibinfo{note}{{Oxford, UK: Univ. Pr. (2007) 710 p}}.

\bibitem{Beltran:2008xg}
\bibinfo{author}{\bibfnamefont{M.}~\bibnamefont{Beltran}},
  \bibinfo{author}{\bibfnamefont{D.}~\bibnamefont{Hooper}},
  \bibinfo{author}{\bibfnamefont{E.~W.} \bibnamefont{Kolb}}, \bibnamefont{and}
  \bibinfo{author}{\bibfnamefont{Z.~C.} \bibnamefont{Krusberg}},
  \bibinfo{journal}{Phys. Rev.} \textbf{\bibinfo{volume}{D80}},
  \bibinfo{pages}{043509} (\bibinfo{year}{2009}), \eprint{0808.3384}.

\bibitem{Bell:2008ey}
\bibinfo{author}{\bibfnamefont{N.~F.} \bibnamefont{Bell}},
  \bibinfo{author}{\bibfnamefont{J.~B.} \bibnamefont{Dent}},
  \bibinfo{author}{\bibfnamefont{T.~D.} \bibnamefont{Jacques}},
  \bibnamefont{and} \bibinfo{author}{\bibfnamefont{T.~J.}
  \bibnamefont{Weiler}}, \bibinfo{journal}{Phys. Rev.}
  \textbf{\bibinfo{volume}{D78}}, \bibinfo{pages}{083540}
  (\bibinfo{year}{2008}), \eprint{0805.3423}.

\bibitem{Kachelriess:2007aj}
\bibinfo{author}{\bibfnamefont{M.}~\bibnamefont{Kachelriess}} \bibnamefont{and}
  \bibinfo{author}{\bibfnamefont{P.~D.} \bibnamefont{Serpico}},
  \bibinfo{journal}{Phys. Rev.} \textbf{\bibinfo{volume}{D76}},
  \bibinfo{pages}{063516} (\bibinfo{year}{2007}), \eprint{0707.0209}.

\bibitem{Hooper:2008zg}
\bibinfo{author}{\bibfnamefont{D.}~\bibnamefont{Hooper}},
  \bibinfo{journal}{Phys. Rev.} \textbf{\bibinfo{volume}{D77}},
  \bibinfo{pages}{123523} (\bibinfo{year}{2008}), \eprint{0801.4378}.

\bibitem{Dent:2008qy}
\bibinfo{author}{\bibfnamefont{J.~B.} \bibnamefont{Dent}},
  \bibinfo{author}{\bibfnamefont{R.~J.} \bibnamefont{Scherrer}},
  \bibnamefont{and} \bibinfo{author}{\bibfnamefont{T.~J.}
  \bibnamefont{Weiler}}, \bibinfo{journal}{Phys. Rev.}
  \textbf{\bibinfo{volume}{D78}}, \bibinfo{pages}{063509}
  (\bibinfo{year}{2008}), \eprint{0806.0370}.

\bibitem{Arina:2007tm}
\bibinfo{author}{\bibfnamefont{C.}~\bibnamefont{Arina}} \bibnamefont{and}
  \bibinfo{author}{\bibfnamefont{N.}~\bibnamefont{Fornengo}},
  \bibinfo{journal}{JHEP} \textbf{\bibinfo{volume}{11}}, \bibinfo{pages}{029}
  (\bibinfo{year}{2007}), \eprint{0709.4477}.

\bibitem{Haber:1984rc}
\bibinfo{author}{\bibfnamefont{H.~E.} \bibnamefont{Haber}} \bibnamefont{and}
  \bibinfo{author}{\bibfnamefont{G.~L.} \bibnamefont{Kane}},
  \bibinfo{journal}{Phys. Rept.} \textbf{\bibinfo{volume}{117}},
  \bibinfo{pages}{75} (\bibinfo{year}{1985}).

\bibitem{Martin:1997ns}
\bibinfo{author}{\bibfnamefont{S.~P.} \bibnamefont{Martin}}
  (\bibinfo{year}{1997}), \eprint{hep-ph/9709356}.

\bibitem{Adulpravitchai:2009re}
\bibinfo{author}{\bibfnamefont{A.}~\bibnamefont{Adulpravitchai}},
  \bibinfo{author}{\bibfnamefont{M.}~\bibnamefont{Lindner}},
  \bibinfo{author}{\bibfnamefont{A.}~\bibnamefont{Merle}}, \bibnamefont{and}
  \bibinfo{author}{\bibfnamefont{R.~N.} \bibnamefont{Mohapatra}},
  \bibinfo{journal}{Phys. Lett.} \textbf{\bibinfo{volume}{B680}},
  \bibinfo{pages}{476} (\bibinfo{year}{2009}), \eprint{0908.0470}.

\bibitem{Kayser:1989iu}
\bibinfo{author}{\bibfnamefont{B.}~\bibnamefont{Kayser}},
  \bibinfo{author}{\bibfnamefont{F.}~\bibnamefont{Gibrat-Debu}},
  \bibnamefont{and} \bibinfo{author}{\bibfnamefont{F.}~\bibnamefont{Perrier}},
  \bibinfo{journal}{World Sci. Lect. Notes Phys.}
  \textbf{\bibinfo{volume}{25}}, \bibinfo{pages}{1} (\bibinfo{year}{1989}).

\bibitem{Falk:1994es}
\bibinfo{author}{\bibfnamefont{T.}~\bibnamefont{Falk}},
  \bibinfo{author}{\bibfnamefont{K.~A.} \bibnamefont{Olive}}, \bibnamefont{and}
  \bibinfo{author}{\bibfnamefont{M.}~\bibnamefont{Srednicki}},
  \bibinfo{journal}{Phys. Lett.} \textbf{\bibinfo{volume}{B339}},
  \bibinfo{pages}{248} (\bibinfo{year}{1994}), \eprint{hep-ph/9409270}.

\bibitem{Amsler:2008zzb}
\bibinfo{author}{\bibfnamefont{C.}~\bibnamefont{Amsler}} \emph{et~al.},
  \bibinfo{journal}{Phys. Lett.} \textbf{\bibinfo{volume}{B667}},
  \bibinfo{pages}{1} (\bibinfo{year}{2008}).

\bibitem{Bellgardt:1987du}
\bibinfo{author}{\bibfnamefont{U.}~\bibnamefont{Bellgardt}} \emph{et~al.},
  \bibinfo{journal}{Nucl. Phys.} \textbf{\bibinfo{volume}{B299}},
  \bibinfo{pages}{1} (\bibinfo{year}{1988}).

\bibitem{Abe:2007ev}
\bibinfo{author}{\bibfnamefont{K.}~\bibnamefont{Abe}} \emph{et~al.}
  (\bibinfo{year}{2007}), \eprint{0708.3272}.

\bibitem{Fayet:2007ua}
\bibinfo{author}{\bibfnamefont{P.}~\bibnamefont{Fayet}},
  \bibinfo{journal}{Phys. Rev.} \textbf{\bibinfo{volume}{D75}},
  \bibinfo{pages}{115017} (\bibinfo{year}{2007}), \eprint{hep-ph/0702176}.

\bibitem{Raspereza:2002ni}
\bibinfo{author}{\bibfnamefont{A.}~\bibnamefont{Raspereza}}
  (\bibinfo{year}{2002}), \eprint{hep-ex/0209021}.

\bibitem{Erkoca:2010qx}
\bibinfo{author}{\bibfnamefont{A.~E.} \bibnamefont{Erkoca}},
  \bibinfo{author}{\bibfnamefont{G.}~\bibnamefont{Gelmini}},
  \bibinfo{author}{\bibfnamefont{M.~H.} \bibnamefont{Reno}}, \bibnamefont{and}
  \bibinfo{author}{\bibfnamefont{I.}~\bibnamefont{Sarcevic}}
  (\bibinfo{year}{2010}), \eprint{1002.2220}.

\bibitem{Rott:2009hr}
\bibinfo{author}{\bibfnamefont{C.}~\bibnamefont{Rott}} \bibnamefont{and}
  \bibinfo{author}{\bibfnamefont{f.~t.~I.} \bibnamefont{Collaboration}}
  (\bibinfo{year}{2009}), \eprint{0912.5183}.

\bibitem{Galli:2009zc}
\bibinfo{author}{\bibfnamefont{S.}~\bibnamefont{Galli}},
  \bibinfo{author}{\bibfnamefont{F.}~\bibnamefont{Iocco}},
  \bibinfo{author}{\bibfnamefont{G.}~\bibnamefont{Bertone}}, \bibnamefont{and}
  \bibinfo{author}{\bibfnamefont{A.}~\bibnamefont{Melchiorri}},
  \bibinfo{journal}{Phys. Rev.} \textbf{\bibinfo{volume}{D80}},
  \bibinfo{pages}{023505} (\bibinfo{year}{2009}), \eprint{0905.0003}.

\bibitem{Slatyer:2009yq}
\bibinfo{author}{\bibfnamefont{T.~R.} \bibnamefont{Slatyer}},
  \bibinfo{author}{\bibfnamefont{N.}~\bibnamefont{Padmanabhan}},
  \bibnamefont{and} \bibinfo{author}{\bibfnamefont{D.~P.}
  \bibnamefont{Finkbeiner}}, \bibinfo{journal}{Phys. Rev.}
  \textbf{\bibinfo{volume}{D80}}, \bibinfo{pages}{043526}
  (\bibinfo{year}{2009}), \eprint{0906.1197}.

\bibitem{Barger:2007hj}
\bibinfo{author}{\bibfnamefont{V.~D.} \bibnamefont{Barger}},
  \bibinfo{author}{\bibfnamefont{W.-Y.} \bibnamefont{Keung}}, \bibnamefont{and}
  \bibinfo{author}{\bibfnamefont{G.}~\bibnamefont{Shaughnessy}},
  \bibinfo{journal}{Phys. Lett.} \textbf{\bibinfo{volume}{B664}},
  \bibinfo{pages}{190} (\bibinfo{year}{2008}), \eprint{0709.3301}.

\bibitem{Delaunay:2008pc}
\bibinfo{author}{\bibfnamefont{C.}~\bibnamefont{Delaunay}},
  \bibinfo{author}{\bibfnamefont{P.~J.} \bibnamefont{Fox}}, \bibnamefont{and}
  \bibinfo{author}{\bibfnamefont{G.}~\bibnamefont{Perez}}
  (\bibinfo{year}{2008}), \eprint{0812.3331}.

\bibitem{Yuksel:2007ac}
\bibinfo{author}{\bibfnamefont{H.}~\bibnamefont{Yuksel}},
  \bibinfo{author}{\bibfnamefont{S.}~\bibnamefont{Horiuchi}},
  \bibinfo{author}{\bibfnamefont{J.~F.} \bibnamefont{Beacom}},
  \bibnamefont{and} \bibinfo{author}{\bibfnamefont{S.}~\bibnamefont{Ando}},
  \bibinfo{journal}{Phys. Rev.} \textbf{\bibinfo{volume}{D76}},
  \bibinfo{pages}{123506} (\bibinfo{year}{2007}), \eprint{0707.0196}.

\bibitem{Bergstrom:1997fj}
\bibinfo{author}{\bibfnamefont{L.}~\bibnamefont{Bergstrom}},
  \bibinfo{author}{\bibfnamefont{P.}~\bibnamefont{Ullio}}, \bibnamefont{and}
  \bibinfo{author}{\bibfnamefont{J.~H.} \bibnamefont{Buckley}},
  \bibinfo{journal}{Astropart. Phys.} \textbf{\bibinfo{volume}{9}},
  \bibinfo{pages}{137} (\bibinfo{year}{1998}), \eprint{astro-ph/9712318}.

\bibitem{Ahmed:2001eh}
\bibinfo{author}{\bibfnamefont{M.}~\bibnamefont{Ahmed}} \emph{et~al.},
  \bibinfo{journal}{Phys. Rev.} \textbf{\bibinfo{volume}{D65}},
  \bibinfo{pages}{112002} (\bibinfo{year}{2002}), \eprint{hep-ex/0111030}.

\bibitem{Lavoura:2003xp}
\bibinfo{author}{\bibfnamefont{L.}~\bibnamefont{Lavoura}},
  \bibinfo{journal}{Eur. Phys. J.} \textbf{\bibinfo{volume}{C29}},
  \bibinfo{pages}{191} (\bibinfo{year}{2003}), \eprint{hep-ph/0302221}.

\bibitem{Raidal:2008jk}
\bibinfo{author}{\bibfnamefont{M.}~\bibnamefont{Raidal}} \emph{et~al.},
  \bibinfo{journal}{Eur. Phys. J.} \textbf{\bibinfo{volume}{C57}},
  \bibinfo{pages}{13} (\bibinfo{year}{2008}), \eprint{0801.1826}.

\bibitem{Aubert:2005wa}
\bibinfo{author}{\bibfnamefont{B.}~\bibnamefont{Aubert}} \emph{et~al.},
  \bibinfo{journal}{Phys. Rev. Lett.} \textbf{\bibinfo{volume}{96}},
  \bibinfo{pages}{041801} (\bibinfo{year}{2006}), \eprint{hep-ex/0508012}.

\bibitem{Dreiner:2008tw}
\bibinfo{author}{\bibfnamefont{H.~K.} \bibnamefont{Dreiner}},
  \bibinfo{author}{\bibfnamefont{H.~E.} \bibnamefont{Haber}}, \bibnamefont{and}
  \bibinfo{author}{\bibfnamefont{S.~P.} \bibnamefont{Martin}}
  (\bibinfo{year}{2008}), \eprint{0812.1594}.

\bibitem{Mertig:1990an}
\bibinfo{author}{\bibfnamefont{R.}~\bibnamefont{Mertig}},
  \bibinfo{author}{\bibfnamefont{M.}~\bibnamefont{Bohm}}, \bibnamefont{and}
  \bibinfo{author}{\bibfnamefont{A.}~\bibnamefont{Denner}},
  \bibinfo{journal}{Comput. Phys. Commun.} \textbf{\bibinfo{volume}{64}},
  \bibinfo{pages}{345} (\bibinfo{year}{1991}).

\bibitem{Kayser:1982br}
\bibinfo{author}{\bibfnamefont{B.}~\bibnamefont{Kayser}},
  \bibinfo{journal}{Phys. Rev.} \textbf{\bibinfo{volume}{D26}},
  \bibinfo{pages}{1662} (\bibinfo{year}{1982}).

\end{thebibliography}

\end{document}